\documentclass[11pt]{article}
\usepackage{amstext}
\usepackage{amssymb}
\usepackage{epsf}

\newcommand{\unity}{{\mathbf 1}}
\newcommand{\be}{\begin{equation}}
\newcommand{\ee}{\end{equation}}
\newcommand{\bea}{\begin{eqnarray}}
\newcommand{\eea}{\end{eqnarray}}
\newcommand{\bref}[1]{(\ref{#1})}

\newcommand{\veps}{\varepsilon}
\newcommand{\ct}[1]{\cite{#1}}

\newcommand{\lder}[2]{\frac{\partial_l#1}{\partial #2}}
\newcommand{\rder}[2]{\frac{\partial_r#1}{\partial #2}}
\newcommand{\gh}[1]{{\cal #1}}

\makeatletter

\def\restric#1#2{{\left. #1 \right|_{#2}}}
\def\dif{{\rm d}}
\def\deriv{\@ifnextchar[{\@deriv}{\@deriv[]}}
   \def\@deriv[#1]#2#3{\mathchoice%
{{\dif^{#1}#2\over\dif{#3}^{#1}}}{{\dif^{#1}#2/\dif{#3}^{#1}}}%
{{\dif^{#1}#2\over\dif{#3}^{#1}}}{{\dif^{#1}#2/\dif{#3}^{#1}}}}

\def\secteqno{\@addtoreset{equation}{section}%
\def\theequation{\thesection.\arabic{equation}}}
\def\endsecteqno{\def\theequation{\@ifundefined{chapter}%
{\arabic{equation}}{\thechapter.\arabic{equation}}}}
\newcounter{subequation}
\def\thesubequation{\alph{subequation}}
\def\sneqnarray{\stepcounter{equation}\let\@currentlabel=\theequation
\setcounter{subequation}{1}
\def\@eqnnum{{\rm (\theequation.\thesubequation)}}
\global\@eqcnt\z@\tabskip\@centering\let\\=\@eqncr\let\@@eqncr=\@@sneqncr
$$\halign to \displaywidth\bgroup\@eqnsel\hskip\@centering
 $\displaystyle\tabskip\z@{##}$&\global\@eqcnt\@ne
 \hskip 2\arraycolsep \hfil${##}$\hfil
 &\global\@eqcnt\tw@ \hskip 2\arraycolsep $\displaystyle\tabskip\z@{##}$\hfil
  \tabskip\@centering&\llap{##}\tabskip\z@\cr}
\def\endsneqnarray{\@@sneqncr\egroup $$\global\@ignoretrue}
\def\@@sneqncr{\let\@tempa\relax
   \ifcase\@eqcnt \def\@tempa{& & &}\or \def\@tempa{& &}
   \else \def\@tempa{&}\fi
     \@tempa \if@eqnsw\@eqnnum\stepcounter{subequation}\fi
     \global\@eqnswtrue\global\@eqcnt\z@\cr}

\def\nobiblabels{\def\@lbibitem[##1]##2{\@bibitem{##2}}}

\def\tr{\mathop{\rm tr}\nolimits}

\makeatother

\def\AP#1#2#3{ {{\sl Ann.\,Phys.\,}(N.Y.)\,}
    {\bf  {#1}} ({#2}) {#3}}

\def\CCG#1#2#3{ {\sl Class.\,Quantum\,Grav.\,}
    {\bf  {#1}} ({#2}) {#3}}
\def\IJMPA#1#2#3{ {\sl Int.\,J.\,Mod.\,Phys.\,}
    {\bf A{#1}} ({#2}) {#3}}

\def\NC#1#2#3{ {\sl Nuovo\,Cim.\,}
    {\bf  {#1}} ({#2}) {#3}}
\def\NPB#1#2#3{ {\sl Nucl.\,Phys.\,}
    {\bf B{#1}} ({#2}) {#3}}

\def\PLB#1#2#3{ {\sl Phys.\,Lett.\,}
    {\bf B{#1}} ({#2}) {#3}}
\def\PR#1#2#3{ {\sl Phys.\,Rep.\,}
    {\bf  {#1}} ({#2}) {#3}}

\def\PRD#1#2#3{ {\sl Phys.\,Rev.\,}
    {\bf D{#1}} ({#2}) {#3}}

\secteqno

\begin{document}

\begin{titlepage}

\begin{flushright}

KUL-TF-96/5\\
hep-th/9607215\\
July 1996\\

\end{flushright}

\vspace{5mm}

\begin{center}

{\LARGE\bf Higher Loop Anomalies and their Consistency Conditions
           in Nonlocal Regularization}

\vskip 2.cm

{\sc Jordi Par\'{\i}s$^\sharp$ and Walter Troost$^\flat$}

\vskip 0.5cm

\small{\it{Instituut voor Theoretische Fysica}}\\
\small{\it{Katholieke Universiteit Leuven}}\\
\small{\it{Celestijnenlaan 200D}}\\
\small{\it{B-3001 Leuven, Belgium}}\\
[2.5cm]

{\bf Abstract}

\end{center}

\begin{quote}
An algebraic program of computation and characterization of higher loop
BRST anomalies is presented.
We propose a procedure for disentangling a genuine
{\it local} higher loop anomaly from
the quantum dressings of lower loop anomalies.
For such higher loop anomalies we derive a local consistency condition,
which is the generalisation of the Wess-Zumino condition
for the one-loop anomaly.
The development is presented in the framework of the
field-antifield formalism, making use of a nonlocal regularization method.
The theoretical construction is exemplified by explicitly computing
the two-loop anomaly of chiral $W_3$ gravity. We also give, for the first
time, an explicit check of the local two--loop consistency condition
that is associated with this anomaly.

\vspace{15mm}

\hrule width 5.cm

{\small \noindent $^\flat$ Onderzoeksleider N.F.W.O., Belgium\\
        \hspace*{4pt} E-mail: Walter.Troost@fys.kuleuven.ac.be\\
        \noindent $^\sharp$ E-mail: Jordi.Paris@fys.kuleuven.ac.be}

\normalsize

\end{quote}

\end{titlepage}

\section{Introduction}

\hspace{\parindent}%
The phenomenon of gauge anomalies \ct{abbj} --obstructions to the
implementation of classical gauge symmetries at quantum level-- has been
known for quite a long time to deeply relate to the algebraic structure of
gauge theories. Its algebraic analysis, initiated in
the pioneering work of Becchi, Rouet and Stora \cite{brs}, has since then
been a fruitful tool to determine the viability of the renormalization
program for such type of theories \cite{Piguet}.
The BRS approach defines an anomaly as an
obstruction to the fulfilment of the BRST-Ward identities.
At lowest order their characterization is reduced to the algebraic
problem of classifying solutions of the Wess-Zumino consistency conditions
\ct{wz}. Anomaly candidates appear as local cohomology classes
at ghost number one of the nilpotent operator $\delta$ generating the BRST
symmetry \ct{brs,tyu} of the classical theory.

BRS ideas have later been generalized and further extended
by means of (among other approaches) the so-called
antibracket-antifield formalism,
or Batalin-Vilkovisky formalism \cite{bv81} or, in short,
FA formalism (see \cite{rev1,rev2,bvboek} for recent reviews).
This proposal starts by first organizing
the fields required to covariantly quantize the theory and their
corresponding BRS sources \ct{Zinn}
in a set of fields and antifields $\{\Phi^A,\,\Phi^*_A\}$ viewed as
local coordinates of an extended ``phase space'' $\gh M$, which is
endowed with an odd symplectic
structure $(\cdot,\cdot)$ --the antibracket--
\be
 (X,Y) = \left(\frac{\partial_r X}{\partial \Phi^A}
         \frac{\partial_l Y}{\partial \Phi^*_A}-
         \frac{\partial_r X}{\partial \Phi^*_A}
         \frac{\partial_l Y}{\partial \Phi^A}\right),
\label{antibracket}
\ee
and, for the quantum theory,
with a nilpotent second order differential operator $\Delta$
$$
   \Delta\equiv (-1)^{(A+1)}
   \frac{\partial_r}{\partial\Phi^A}
   \frac{\partial_r}{\partial\Phi^*_A},
   \quad\quad \Delta^2\equiv0.
$$

In this geometrical setting, the classical BRST structure of a given gauge
theory is encoded in a generalized action $S(\Phi,\Phi^*)$.
Its antifield expansion
\be
  S(\Phi,\Phi^*)=\gh S(\Phi)+ \sum_{n\geq 1}
  \frac1{n!}\Phi^*_{A_1}\ldots\Phi^*_{A_n} R^{A_n\ldots A_1}(\Phi),
\label{gfps}
\ee
contains as coefficients a suitable gauge-fixed action, ${\gh S}(\Phi)$,
the BRST symmetry transformations $\delta\Phi^A=R^A(\Phi)$, and
higher order structure functions, $R^{A_n\ldots A_1}(\Phi)$
describing the classical BRST structure.
The basic equation  is the so-called classical master equation
\be
   (S,S)=0,
\label{cme}
\ee
of which $S$ is a  proper solution. Due to this equation,
$S$ itself is invariant under the generalized BRST symmetry
$\hat\delta$ in $\gh M$ defined by
\be
   \hat\delta F(\Phi,\Phi^*)\equiv (F,S).
\label{gen brst}
\ee

Quantization of the theory is accomplished by path--integrating
over a suitable quantum extension $W$ of the action $S$ \bref{gfps}
\be
   W= S+\sum_{p\geq 1}{\hbar}^p M_p,
\label{quantum action}
\ee
where we anticipated  the presence of local counterterms to guarantee
finiteness of the theory (and possibly to preserve quantum BRST
invariance as well).
The corresponding effective action $\Gamma(\Phi,\Phi^*)$
incorporates then all loop effects to the classical BRST structure
functions and is thus interpreted, in analogy with $S$ \bref{gfps}, as the
generating functional of their quantum counterparts. In this way the quantum
deformation of the classical BRST structure is naturally described by
means of the quantum analog of the classical master equation \bref{cme}
--the (anomalous) Zinn-Justin equation \ct{Zinn}--
\be
   \frac12(\Gamma,\Gamma)=-i\hbar(\gh A \cdot\Gamma),
\label{ward identity}
\ee
where $(\gh A \cdot\Gamma)$ stands for
the generating functional of 1PI Green functions with one insertion of
a composite field $\gh A$. This composite operator $\gh A$
parametrizes the departure from the classical BRST structure
due to quantum corrections, and is interpreted as the BRST anomaly.
The FA formalism gives an explicit expression for this anomaly
\be
   \gh A =
   \left[\Delta W+\frac{i}{2\hbar}(W,W)\right],
\label{anomaly}
\ee
which however is very formal, since the action of the operator $\Delta$
on {\em local} expressions is ill-defined, proportional to $\delta(0)$.
Whereas it is possible to
obtain a corresponding expression in renormalised perturbation theory
using Zimmerman's normal products \ct{BPHZBV}, we will follow
in this paper a different
method where, on the regularised level (i.e. also before the actual
renormalisation) these expressions do make sense because effectively all
quantities on which $\Delta$ acts will be {\em nonlocal}.

On the level of the effective action the generalized BRST transformation
$\hat\delta$ \bref{gen brst} is \ct{rev2}
\be
   \hat\delta_Q\, (\gh B\cdot\Gamma)\equiv
   \left((\gh B\cdot\Gamma),\Gamma\right)=
   \left((\sigma \gh B+ \gh B\,\gh A)\cdot\Gamma\right),
\label{eff brst 2}
\ee
relation which extends to the general anomalous case
the result previously obtained in \ct{anselmi}
for nonanomalous theories, namely
$\left((\gh B\cdot\Gamma),\Gamma\right)=
\left((\sigma \gh B)\cdot\Gamma\right)$.
In these expressions, the so-called {\it quantum BRST operator} $\sigma$,
a possible%
\footnote{The alternative is to include the $\gh A$-term: then the
transformation is nilpotent \cite{bvboek}.}
quantum generalization \ct{lt85a}
of the classical BRST transformation \bref{gen brst}, is
formally given in terms of the generalized quantum action $W$
\bref{quantum action} by
\be
   \sigma \gh B\equiv (\gh B, W)-i\hbar \Delta \gh B.
\label{qbo}
\ee

Equation \bref{eff brst 2}, together with the FA form of the BRST anomaly
\bref{anomaly}, is one of the cornerstones of the treatment of anomalies
in this framework. The operator $\sigma$ \bref{qbo} contained in it
determines a set of local consistency conditions for $\gh A$
in \bref{ward identity}. Indeed, a straightforward
use of the Jacobi identity for the antibracket leads to the following
equation for $(\gh A\cdot\Gamma)$:
\be
    \left((\Gamma,\Gamma),\Gamma\right)\equiv 0 \,\Rightarrow\,
    \left((\gh A\cdot\Gamma),\Gamma\right)\equiv 0,
\label{nlcc}
\ee
and therefore, by \bref{eff brst 2},
\be \sigma\gh A=0 \, .\label{lcc} \ee

The vanishing of the action of the quantum BRST operator $\sigma$ on
$\gh A$ is then the formal expression for the
consistency condition on the BRST anomaly. If we expand it in powers of
$\hbar$ (i.e. a loop-wise expansion),
$\gh A=\sum_{p=1}\hbar^{p-1} \gh A_{p}$,
we obtain the formal expression for
the $p$-loop consistency conditions
\bea
     &&(\gh A_1, S)=0,
\label{oneloop}\\
     &&(\gh A_2, S) +(\gh A_1,M_1) -i\Delta \gh A_1=0,
\label{twoloop}\\
     &&(\gh A_p, S) +\sum^{p-1}_{q=1} (\gh A_q,M_{p-q})
     -i\Delta \gh A_{p-1}=0,
     \quad\quad p > 2.
\label{nloop}
\eea
The first of these equations is clearly the Wess-Zumino consistency
condition for the lowest order part of the anomaly \ct{anombv,hlw90}. The
rest of the relations, \bref{twoloop} and \bref{nloop}, are the
generalization for higher loop anomalies.

The characterization of one loop anomalies by means of \bref{oneloop} and
the associated cohomological problem, and its calculation through
\bref{anomaly}, have received considerable attention in
recent years.
A similar algebraic higher loop analysis, on the other hand,
is almost completely missing. In no small measure, this is due to the fact
that, although the Field-Antifield
formalism computes and characterizes the BRST anomalies from an algebraic
point of view in a straightforward manner, the resulting equations remain
formal since they involve the action of $\Delta$ on local expressions:
the equation \bref{oneloop} for the one loop part is the only exception to
this. In ref.\,\ct{white92} the
cohomological question was adressed using condition
\bref{nlcc}, and seemed to lead to the conclusion that all higher loop
anomalies would satisfy the Wess-Zumino consistency condition
\bref{oneloop}. This is in disagreement with the formal
FA result \bref{twoloop} and \bref{nloop}, which indicates the presence of
extra pieces related with the action of $\Delta$ on lower order anomalies%
\footnote{In this respect, see ref.\,\ct{watts}, where generalized
consistency conditions for pure $W_3$ gravity are obtained by algebraic
arguments, not relying on perturbation theory.}.
In \ct{BPHZBV}, an all-loop perturbative treatment was set up using
well-defined expressions in terms of normal product operators, but the
issue of consistency conditions was not addressed.
Whereas this treatment showed that it is feasable to set up the quantum
theory without introducing the $\Delta$ operator, it remains nevertheless
true that, formally, this operator is very useful to highlight the
structure of the FA quantum theory. On the one-loop level, a
Pauli-Villars scheme has been introduced \ct{anombv} to deal in
particular with the action of $\Delta$ on a local $W$
in eq.\,\bref{anomaly}, avoiding the $\delta(0)$ type singularity, and
even with its action on a generic local functional or composite field
\ct{frank}.

In this paper we use the nonlocal regularization method of \ct{emkw91},
which was shown in \ct{p95}
to deal very effectively with $\Delta$, by making sure that it acts
only on {\em non-local} expressions at the regularized level,
thus avoiding problems with its definition while keeping it as a powerful
tool to investigate the quantum theory. The treatment in \ct{p95} fell
short of a treatment of higher loop anomalies however, in particular of
the consistency conditions that they satisfy.

Our main purpose in this paper is therefore to push the computation and
algebraic characterization of higher loop anomalies  further.
We will show how the regularisation problems of the higher loop
consistency equations are overcome
in the framework of the nonlocally regularized FA formalism.
First we demonstrate how information about genuine, local higher
loop anomalies can be extracted from the nonlocally
regulated version of eq.\,\bref{anomaly}.
Actually, they arise from quantum dressings of terms contained in the
nonlocally regulated version of the insertion $\gh A$ \bref{anomaly} that
are evanescent: they formally disappear when the
non-local cutoff is removed but may nevertheless give a finite contribution
when the limiting process is treated more accurately.
Once their origin is identified, the nonlocally regulated version
of the BRST Ward identity \bref{ward identity} and of its associated
Jacobi identity \bref{nlcc} lead to a set of local consistency
conditions for them.

We have organized the paper as follows.
After summarizing in section \ref{nrfa} the basics of the
nonlocally regulated field-antifield formalism as developed in \ct{p95},
we show and illustrate in section \ref{hla}  how to obtain the local
higher loop anomalies, and in section \ref{hlcc} their consistency
conditions. The theoretical part is further supported
in section \ref{w3example} by explicitly computing the two-loop anomaly
in the example of chiral $W_3$ gravity and  presenting,
for the first time, an explicit
and complete check of the local two--loop consistency condition
associated with this anomaly.
Section \ref{conclusions} summarizes conclusions and perspectives.

\section{Nonlocal Regularization in the FA framework}
\label{nrfa}

\hspace{\parindent}%
The nonlocally regularized FA formalism \ct{p95} arises as an extension
and reformulation of the nonlocal regularization method of
\mbox{refs.\,\ct{emkw91}}
along the line of the antibracket--antifield formalism,
giving sense to both diagramatic and formal algebraic computations in
this framework. Roughly speaking, nonlocal regularization starts by
reformulating the theory in an alternative but equivalent nonlocal way,
in which auxiliary fields completely encode in their loops the
divergencies of the original theory. Eliminating these loops, through the
elimination of these auxiliary fields by putting them equal to their
on--shell value, gets rid of their divergent loops and
regularizes the theory. Remarkably, this can be done while preserving
the BRST (or gauge) symmetry: the resulting nonlocal, regularized theory,
is still invariant classically under a nonlocal, distorted version of the
original BRST transformation. Previously ill--defined quantities and
manipulations --ill--defined because of the locality of the expressions--
such as the expression for the FA BRST anomaly \bref{anomaly}
or its consistency condition \bref{lcc},
acquire a well--defined meaning in this process.
Of course, as always, after the regularization step one still has to take
into account the renormalization to arrive at the end results.
We provide for this by incorporating the inclusion of
the necessary counterterms from the start.

In this section  we summarize the main ideas of this
regularized framework, paving the way for our basic goal, namely to
identify the origin of higher loop anomalies in this approach
and to explore their consistency conditions.

Consider a FA quantized gauge theory and decompose the proper
solution $S$ \bref{gfps}  into free and interacting parts
\be
  S(\Phi,\Phi^*)= F(\Phi)+ I_{\text{cl}}(\Phi,\Phi^*),
  \quad\quad \mbox{\rm with}\quad\quad
   F(\Phi)=\frac12\Phi^A  F_{AB} \Phi^B,
\label{original action}
\ee
and where%
\footnote{\label{notation}
We use de Witt's notation, where the space--time point on
which a field depends is included in the index of that field.
Summation over this index includes a space--time integral.
Functional derivatives with respect to fields or antifields
are indicated with a lower or upper index respectively: for example,
$W^A_{\,B}=\frac{\partial_r\partial_l W}{\partial\Phi^B\partial\Phi^*_A}$
and
$W_{A\,B}=\frac{\partial_r\partial_l W}{\partial\Phi^B\partial\Phi^A}$.
This notation will  apply throughout this paper,
for example for $F_{AB}$ in \bref{original action}, or
later in this section in \bref{quantum delta} and \bref{inverse k},
except if the indexed object is defined explicitly, as in the following
formulas for $\gh R$, $\veps$ and $\gh O$.}
the classical interaction part $I_{\text{cl}}(\Phi,\Phi^*)$
is assumed to
be analytic in $\Phi^A$ around $\Phi^A=0$. The
quantum action \bref{quantum action} has a similar perturbative expansion
$$
   W= F+I_{\text{cl}}+\sum_{p\geq 1}{\hbar}^p M_p\equiv F+I,
$$
in terms of a generalized quantum interaction $I(\Phi,\Phi^*)$
which includes  the $\Lambda$-dependent counterterms mentioned above.

The cut--off parameter $\Lambda^2$ and a smearing operator $\veps$ are
introduced as follows. First choose a field independent%
\footnote{One might relax this restriction if the need arises, but for
simplicity we will not do so.}
(graded symmetric) operator $(T^{-1})^{AB}$ in
such a way that a second order derivative ``regulator'' $\gh R^A_{\,B}$
arises through the combination
\be
   \gh R^A_{\,B}=(T^{-1})^{AC} F_{CB}.
\label{regulator}
\ee
The regulator is intended to provide a momentum cutoff through
an ubiquitous smearing operator $\gh\veps^A_{\,B}$:
\be
   \gh\veps^A_{\,B}= \exp\left(\frac{\gh R^A_{\,B}}{2\Lambda^2}\right).
\label{smearing op}
\ee

The original phase space $\gh M$ is now temporarily enlarged
with the so-called ``shadow''
fields and antifields $\{\Psi^A,\Psi^*_A\}$,
which have the same statistics as
the original fields $\{\Phi^A,\Phi^*_A\}$,  extending
also the antibracket structure
\bref{antibracket} in the natural way.
For these fields one takes as (minus) the propagator
the ''shadow propagator'' $\gh O^{AB}$:
\be
   \gh O^{AB}= \left(\frac{(\gh\veps^2-1)}{ F}\right)^{AB}=
   \left[\int^1_0\frac{\dif t}{\Lambda^2}\,
   \exp\left(t\frac{\gh R^A_{\,C}}{\Lambda^2}\right)\right](T^{-1})^{CB}.
\label{shadow propagator}
\ee
The interactions are most  conveniently described in terms of a different
set of canonical coordinates
$\{\Theta^A,\Theta^*_A;\Sigma^A,\Sigma^*_A\}$,
related to the original ones by the linear canonical transformation
\bea
  \Theta^A&=& \Phi^A+\Psi^A,\quad\quad
  \Theta^*_A= \left[\Phi^*_B(\gh\veps^2)^B_{\,A}+
  \Psi^*_B (1-\gh\veps^2)^B_{\,A}\right],
\nonumber\\
  \Sigma^A&=&
  \left[(1-\gh\veps^2)^A_{\,B}\Phi^B-(\gh\veps^2)^A_{\,B}\Psi^B\right],
  \quad\quad
  \Sigma^*_A= \Phi^*_A-\Psi^*_A.
\label{new variables}
\eea
The action is now rewritten as follows: replace in the old free action
the original fields $\Phi^A$ with the smeared fields
$(\gh\veps^{-1})^A_{\, B}\Phi^B$, add for the shadow fields a free part
with the propagator \bref{shadow propagator}
constructed above, and for the interaction terms take
the old interaction functional $I$ but change the value of its arguments
from $\{\Phi^A, \Phi^*_A\}$ to $\{\Theta^A, \Theta^*_A\}$
\bref{new variables}. This yields the auxiliary quantum action
\bea
  \tilde W(\Phi,\Phi^*,\Psi,\Psi^*)
    &=& F(\veps^{-1}\Phi)-\frac12\Psi^A \gh O^{-1}_{AB} \Psi^B+
        I(\Theta,\Theta^*)\nonumber\\
    &=& W(\Theta,\Theta^*)+ \frac12\Sigma^A
  \left[\frac{\gh F}{\veps^2}+ \frac{\gh F}{(1-\veps^2)}\right]_{AB}
  \Sigma^B.
\label{aux quantum action}
\eea
The remarkable result of \ct{p95} is that this process does not interfere
with the FA structure, and a fortiori not with the BRST structure of the
theory. This  clarifies the fact that the gauge symmetry is preserved
on the classical level, albeit in a deformed way,
which in fact was the principal motivation for the introduction of this
variant of non-local regularisation \ct{emkw91} for Yang-Mills theories.
In view of the generality of the FA formalism,
it also immediately generalises this property to arbitrary gauge theories
(open algebras, reducible symmetries, etc.) within the realm of the FA
formalism.

All in all, the net result of this preliminary process is an auxiliary
perturbative theory completely equivalent to the original one%
\footnote{For a proof of this equivalence, notice that the sum of the
propagators of the $\Phi$ and $\Psi$ fields in \bref{aux quantum action}
is equal to the original $\Phi$ propagator, and see \cite{zjbook},
section 7.1.}
when no external $\Psi$ lines are considered.
However, the description of the
theory by means of the auxiliary action \bref{aux quantum action}
concentrates the loop divergencies solely in loops of the
auxiliary fields. In this way, regularization of the theory is
achieved  by eliminating the closed loops formed with ``shadow''
lines by hand. This can  be implemented by
putting the auxiliary fields $\Psi$ equal to their  on-shell
values and their antifields to zero%
\footnote{Even though this is not a canonical transformation,
since $\{\Psi,\Psi^*\}$ is not a trivial system, nevertheless
this substitution does not invalidate the FA structure since the master
equation remains valid \ct{p95}.}.
The shadow field equations of motion
\be
   \rder{\tilde{W}(\Phi,\Phi^*;\Psi, \Psi^*=0)}{\Psi^A}=0
  \quad\text{i.e.}\quad
  \Psi^A=\left(\rder{I}{\Phi^B}(\Phi+\Psi,\Phi^*\gh\veps^2)
  \right)\gh O^{BA},
\label{q equations motion}
\ee
are to be solved for $\Psi$, perturbatively both in $\hbar$ and in the
coupling constants, and its solution $\bar\Psi_q(\Phi,\Phi^*)$ substituted
in the auxiliary action \bref{aux quantum action}. The result of this
second step is the final form of the nonlocal quantum action to be used
in (now regularized) perturbative computations:
\be
  W_\Lambda(\Phi,\Phi^*)\equiv\tilde W(\Phi,\Phi^*,\bar\Psi_q,\Psi^*=0).
\label{w lambda}
\ee
Its loopwise expansion
$$
  W_\Lambda= S_\Lambda+\sum_{n\geq 1}\hbar^p M_{p,\Lambda},
$$
provides the nonlocal regularized versions $S_\Lambda$, $M_{p,\Lambda}$,
of the classical action \bref{gfps} and of the
counterterms $M_p$ in \bref{quantum action}, respectively.

More generically, to write down the smeared nonlocal functional
to be used instead of a given classical functional
$\gh F(\Phi,\Phi^*)$, one performs the substitution
\be
   (\Phi,\Phi^*)\longrightarrow (\Phi+\bar\Psi_q,\Phi^*\gh\veps^2).
\label{nlchange}
\ee
The resulting functional
$\gh F_{R}(\Phi,\Phi^*)\equiv
\gh F(\Phi+\bar\Psi_q,\Phi^*\gh\veps^2)$
will then be denoted generically%
\footnote{\label{Rnotation}%
We use the notation $\gh F_{\Lambda}$ for the functional
of the original fields and antifields that, in the $\Lambda$-regulated
theory, replaces the functional $\gh F$. This may already contain the
cutoff $\Lambda$ in certain places. In the regulated theory, these
quantities usually appear, in addition,
with the shifted arguments \bref{nlchange}, resulting in
$\gh F_{\Lambda R}(\Phi,\Phi^*)=
\gh F_\Lambda(\Phi+\bar\Psi_q,\Phi^*\gh\veps^2)$.
It is important to keep in mind the distinction between $\gh F_\Lambda$,
$\gh F_{\Lambda R}$, and $\gh F_{R}$.}
with a subscript $R$ (for ''regulated'').
This substitution \bref{nlchange} can be interpreted
as the nonlocal regularization rule for composite fields:
in the (formal) limit $\Lambda \rightarrow \infty$ we have that
$\veps\rightarrow 1$ and $\bar\Psi_q\rightarrow 0$ \ct{emkw91}
(see \bref{q equations motion} and \bref{shadow propagator}).
This prescription was followed, for example, for the interaction terms in
the action. Sometimes the actual
corresponding quantity to be used in the non-locally
regularised theory (which we denote generically by $\gh F_\Lambda$)
is different: an important example is the kinetic term in the action.
In fact, from \bref{aux quantum action} it follows that
$W_\Lambda-W_{R}$ consists of the terms arising from the
$\Sigma$-terms in \bref{aux quantum action} by putting $\Psi=\bar\Psi_q$.

The transition from the formal to the regularized theory
is generally made  by changing the original
quantities ($S$, $W$, $\ldots$) into their nonlocal counterparts
\mbox{($S_\Lambda$, $W_\Lambda$, $\ldots$)},
possibly with shifted arguments.
Most of the formal relations of the
unregularised theory then become true equalities for the regularised theory
without any other changes. For instance, the relations
characterizing algebraically the regulated BRST symmetry at classical
level are now encoded in the set of equations coming from the regulated
classical master equation $(S_\Lambda,S_\Lambda)=0$.
This equation is  verified by construction \ct{p95}. The classical
BRST structure is therefore preserved at the regularized
level. By the same token, at quantum level the BRST structure and its
possible breakdown are described by means of the regulated
counterpart of the BRST Ward identity \bref{ward identity}
\be
   \frac12(\Gamma_\Lambda,\Gamma_\Lambda)=
   -i\hbar(\gh A_{\Lambda R} \cdot\Gamma_\Lambda),
\label{reg ward ident}
\ee
where $\Gamma_\Lambda$ stands for the  effective action (1PI)
associated to the nonlocally regulated quantum action $W_\Lambda$
\bref{w lambda}, and for the insertion we have followed the
notational convention just explained. Indeed, the obstruction
$\gh A_{\Lambda R}$ parametrizing this breakdown is still of  the
form \bref{anomaly} with  $W\rightarrow W_\Lambda$, i.e.
\be
   \gh A_{\Lambda R}(\Phi,\Phi^*)=
   \left[\Delta W_\Lambda
   +\frac{i}{2\hbar}(W_\Lambda,W_\Lambda)\right](\Phi,\Phi^*)=
   \gh A_{\Lambda}(\Phi+\bar\Psi_q,\Phi^*\gh\veps^2),
\label{reg qme}
\ee
where now the action of the operator $\Delta$
is well--defined thanks to the nonlocality of $W_\Lambda$.

At this point we have regular versions
\bref{reg ward ident} and \bref{reg qme}
of the fundamental equations
\bref{ward identity} and \bref{anomaly} at our
disposal. The content of these equations can be explored further
using the particular properties following from the construction
of the action \bref{w lambda}.
Starting with the insertion $\gh A_{\Lambda R}$ \bref{reg qme},
the action of the operator $\Delta$ on $W_\Lambda$
can be explicitly computed. The
derivatives of $W_{\Lambda}$ \bref{w lambda}
and $\tilde{W}$ \bref{aux quantum action} are related by
\be
   \rder{W_{\Lambda}}{\Phi^A}= \restric{\rder{\tilde W}{\Phi^A}}{q},
   \quad\quad
   \lder{W_{\Lambda}}{\Phi^*_A}= \restric{\lder{\tilde W}{\Phi^*_A}}{q},
\label{relations der w}
\ee
where the ``$q$''-restriction means on the surface
$\{\Psi=\bar\Psi_q(\Phi,\Phi^*), \Psi^*=0\}$.
This yields
\be
  \Delta W_\Lambda(\Phi,\Phi^*)=
  \left[ W^A_{\,B}\, (\delta_\Lambda)^B_{\,C}\,(\gh\veps^2)^C_{\,A} \right]
  (\Phi+\bar\Psi_q,\Phi^*\gh\veps^2)
  \equiv\Omega_W(\Phi+\bar\Psi_q,\Phi^*\gh\veps^2),
\label{final delta reg}
\ee
where the matrix $(\delta_\Lambda)^A_{\,B}$ is defined as
\be
  (\delta_\Lambda)^A_{\,B}=
  \frac{\partial_r(\Phi+\bar\Psi_q)^A}{\partial\Phi^B}=
  {\gh K}^{AC}(\gh O^{-1})_{CB}=
  \left(\delta^A_{\,B}-
  \gh O^{AC}I_{CB}\right)^{-1},
\label{quantum delta}
\ee
in terms of an operator ${\gh K}^{AB}$ whose inverse is given by
\be
  {\gh K}^{-1}_{AB}=
  (\gh O^{-1})_{AB}-I_{AB} \, .
\label{inverse k}
\ee
If we compare this expression with the formal computation of  $\Delta W$,
which would give
\be
   \Delta W=
 \frac{\partial_r\partial_l W}{\partial\Phi^B\partial\Phi^*_A}
   \delta^B_{\,A}= W^A_{\,B} \delta^B_{\,A},
\label{original delta w}
\ee
we can view the regularised expression in \bref{final delta reg}
as  resulting from \bref{original delta w} in two steps.
The first is special, and consists of inserting the extra factor
$\delta_\Lambda \veps^2$ (instead of a delta-function) which may be viewed
as smearing out the $\Delta$ operator. This results in a quantity we call
$\Omega_W(\Phi,\Phi^*)$,  which depends on  $\Lambda$ explicitly
through the presence of $\veps^2$ and $\gh O$.
The second is to change the fields and antifields in the argument of the
functional with  the  substitution \bref{nlchange},
the generic regularisation step.

Analogous manipulations allow to rewrite the second term
in \bref{reg qme} as
$$
  (W_{\Lambda}, W_{\Lambda})(\Phi,\Phi^*)=
  \restric{(\tilde W, \tilde W)(\Theta,\Theta^*)}{q}=
  (W,W)(\Phi+\bar\Psi_q,\Phi^*\gh\veps^2),
$$
and expression \bref{reg qme} as
\be
   \gh A_{\Lambda R}(\Phi,\Phi^*)=
   \left[\Omega_W+\frac{i}{2\hbar}(W,W)\right]
   (\Phi+\bar\Psi_q,\Phi^*\gh\veps^2)\equiv
   \gh A_\Lambda (\Phi+\bar\Psi_q,\Phi^*\gh\veps^2).
\label{fin anomaly 1}
\ee

Assuming that we are working with a renormalisable theory,
the validity of the regularization method requires the
existence of suitable counterterms $M_p$ cancelling the possible
divergencies arising, for example,
in the computation of $\Omega_W(\Phi,\Phi^*)$. As stated in the beginning,
these counterterms have been assumed to be present in $W$ from the start.
Taking the formal $\Lambda\rightarrow\infty$ limit of \bref{fin anomaly 1}
then results in a finite local functional $\bar\gh A(\Phi,\Phi^*)$
\be
   \bar\gh A(\Phi,\Phi^*)=
   \lim_{\Lambda^2\rightarrow\infty}
   \gh A_\Lambda (\Phi+\bar\Psi_q,\Phi^*\gh\veps^2)=
   \lim_{\Lambda^2\rightarrow\infty}
   \left[\Omega_W+ \frac{i}{2\hbar}(W,W)\right](\Phi,\Phi^*) .
\label{fin anomaly 2}
\ee
One would now be tempted to consider this local functional as
the BRST anomaly produced by this regularization procedure and,
after the regularisation step $\bar\gh A\rightarrow\bar\gh A_{R}$,
to use it as insertion, instead of $\gh A_{\Lambda R}$
\bref{fin anomaly 1}, in the Zinn-Justin equation \bref{reg ward ident}.
In fact, a cursory glance confirms that the consistent one--loop anomaly is
then correctly reproduced by the tree level part.
However, a closer investigation \ct{p95} reveals that
the contributions to \bref{fin anomaly 2}
of higher order in $\hbar$ are incomplete: they
reproduce correctly only the one--loop
parts of the contributions to the anomaly
generated by the counterterms $M_p$ of \bref{quantum action},
but {\it not} the genuine, local higher loop anomalies.
For this reason it was conjectured in \ct{p95} that the master
equation might be incomplete.

Nevertheless, if the regularisation scheme used here is correct,
the regulated BRST Ward identity \bref{reg ward ident} must
already contain {\it all} the information about higher loop anomalies.
In other words, the quantum dressing of the insertion $\gh A_{\Lambda R}$
in the regulated theory,
\mbox{$(\gh A_{\Lambda R}\cdot\Gamma_\Lambda)$,} should
generate both the (nonlocal) loop corrections to the lower order
anomalies and the local, genuine higher loop anomalies,
which seem to be absent from $\bar\gh A$.
In the next section we therefore analyse in detail the
$\Lambda\rightarrow\infty$ limiting process.

\section{Higher Loop Anomalies in Nonlocal Regularization}
\label{hla}

\hspace{\parindent}%
To investigate this discrepancy, that the genuine higher loop anomalies
do not to show up clearly in \bref{fin anomaly 2},
we now investigate in detail the limiting process by which
a finite local expression is obtained for the anomaly $\gh A$.

The one-loop anomaly functional is computed as \ct{p95}
\be
   \gh A_1=
   \lim_{\Lambda^2\rightarrow\infty} \left[\Omega_S +i(M_1,S)\right],
\label{oneloop anom}
\ee
where $\Omega_S$ is the lowest order part in the loopwise expansion
of $\Omega_W$ \bref{final delta reg}
\be
    \Omega_S=
  \left[S^A_{\,B}(\delta_\Lambda^{(0)})^B_{\,C}(\gh\veps^2)^C_{\,A}\right],
\label{omega 0}
\ee
and $\delta_\Lambda^{(0)}$ is the classical part of
$\delta_\Lambda$ \bref{quantum delta}
\be
  (\delta_\Lambda^{(0)})^A_{\,B}=
  \left(\delta^A_B-\gh O^{AC}(I_{\text{cl}})_{CB}\right)^{-1}=
   \delta^A_{\,B}+\sum_{n\geq 1}
  \left(\gh O^{AC}(I_{\text{cl}})_{CB}\right)^{n}.
\label{delta lambda}
\ee
To compute the limit of $\Omega_S$, it may be necessary to subtract a
''divergent'' part, typically diverging as a power of $\Lambda$:
this is done through a corresponding $M_1$ via the $(M_1,S)$ term in
\bref{oneloop anom}. Then the $\Lambda\rightarrow\infty$ limit can be
taken, and yields a local $\Lambda$-independent functional
of the fields $\Phi$ and $\Phi^*$.

This functional may now subsequently be used as an insertion in
\bref{reg ward ident} to investigate the next order.
Of course one can not just insert $\gh A_1$ as it stands
(which would give divergent results, even at finite $\Lambda$),
but rather the corresponding regularised functional
($\gh A_1\rightarrow \gh A_{1\,R}$, see \bref{nlchange});
this regularised expression is used in a regularised one--loop
computation (using $W_\Lambda$ and $\Gamma_\Lambda$).
The result of this process,
after the $\Lambda\rightarrow\infty$ limit is taken, can be denoted
$(\gh A_1 \cdot \Gamma)$. In this way one finds
\be
  \lim_{\Lambda^2\rightarrow\infty}
  (\gh A_{\Lambda R}\cdot\Gamma_\Lambda)=
  \gh A_1+ \hbar\left[ \gh A_2+
  (\gh A_1\cdot\Gamma_1)\right] + O(\hbar^2),
\label{claim}
\ee
where $(\gh A_1\cdot\Gamma_1)$ is the one--loop piece of
$(\gh A_1 \cdot \Gamma)$, obtained in the way described above,
and $\gh A_2$ is simply the limit of the remaining terms at that order.
Had we used instead the regulated version of
$\bar\gh A$ \bref{fin anomaly 2}, $\bar\gh A_R$, as insertion, the
result would have been
\be
  \lim_{\Lambda^2\rightarrow\infty}
  (\bar\gh A_R\cdot\Gamma_\Lambda)=
  \gh A_1+ \hbar\left[ \bar\gh A_2+
  (\gh A_1\cdot\Gamma_1)\right] + O(\hbar^2),
\label{mistake}
\ee
with $\bar\gh A_2$ encoding only the contributions to the
complete two--loop anomaly $\gh A_2$
generated by the counterterms $M_p$, $p=1,2$, of \bref{quantum action},
but {\it not} the genuine, local two--loop anomaly.

The reason for the discrepancy between \bref{claim} and \bref{mistake}
can now be understood as follows.
Formally the difference
$\gh A_\Lambda(\Phi,\Phi^*)- \bar\gh A(\Phi,\Phi^*)$
vanishes when one takes the $\Lambda\rightarrow\infty$ limit,
typically as an inverse power of $\Lambda$ times a local operator,
and can therefore properly be called ''evanescent''%
\footnote{''likely to vanish''\ct{dictionary}.}.
One might therefore expect that in the construction above the functional
$\gh A_2-\bar\gh A_2$, which corresponds to this difference, would vanish.
However, in the regularised theory, the quantum corrections to some
matrix elements of the local operator
$\gh A_\Lambda- \bar\gh A$ may diverge. More accurately,
since in the present context no divergences are present at
finite $\Lambda$, the loops behave as positive powers for
$\Lambda\rightarrow\infty$. This divergence may conspire with
the formal vanishing  of $\gh A_\Lambda-\bar\gh A$ to give a
{\em finite non-vanishing} result%
\footnote{A similar behaviour was observed by Gervais and Jevicky
          in \ct{gj76}, on implementing point canonical transformations in
          a time discretized version of the quantum mechanical path
          integral.}.
As a consequence, when using $\bar\gh A$ \bref{fin anomaly 2} as insertion
in \bref{mistake}, this nonvanishing contribution will not be included in
the quantum correction to lower order anomaly insertions, and therefore has
to be added separately,
as a new insertion, in order to reproduce some of the effects of the
primary insertion $\gh A_\Lambda$, which otherwise would be lost.
In other words, proceeding in this way, these quantum corrections to
evanescent pieces are {\em effectively} replaced by their effect:
this turns out to be the source of the
{\em genuine} higher loop anomalies in this approach.

The steps traced above to order $\hbar$ can be retraced for the
higher orders as well. 
At each order, loop corrections of the lower orders appear, but also
additional local terms.
It is crucial to realise that the anomaly functional
\be
\gh A=\gh A_1 +\hbar\gh A_2 +\ldots
\label{truean}
\ee
obtained in this way as a power series in $\hbar$
does {\em not} reproduce the functional $\gh A_{\Lambda}$.
Instead, it {\em does} reproduce, by the very definition of the successive
terms, all the effects of the $\gh A_{\Lambda}$ insertion upon taking the
$\Lambda\rightarrow\infty$ limit.
Namely, although $\gh A_{\Lambda}$ and $\gh A$ differ
by terms with inverse powers
of $\Lambda$, the extra local terms present in $\gh A$ and not in
$\gh A_\Lambda$ are adjusted by construction
so that they reproduce the finite effects of this difference: these arise
from loop divergences. In this way,
when using the regulated counterpart $\gh A_R$ of
the anomaly functional $\gh A$ \bref{truean}
as insertion in the regulated theory
$$
   \gh A_R(\Phi, \Phi^*)\equiv
   \left[\gh A_1 + \hbar \gh A_2 + O(\hbar^2)\right]
   (\Phi+\bar\Psi_q,\Phi^*\gh\veps^2),
$$
it correctly reproduces, in the $\Lambda\rightarrow\infty$ limit,
the results of inserting the quantity
$\gh A_\Lambda$ \bref{fin anomaly 1}, i.e.
\be
  \lim_{\Lambda^2\rightarrow\infty}
  (\gh A_{\Lambda R}\cdot\Gamma_\Lambda)=
  \lim_{\Lambda^2\rightarrow\infty}
  (\gh A_R\cdot\Gamma_\Lambda)=
  \gh A_1+ \hbar\left[ \gh A_2+
  (\gh A_1\cdot\Gamma_1)\right] + O(\hbar^2).
\label{claim2}
\ee

For the sake of definiteness, we now derive a closed expression for
the genuine two-loop anomaly: it will be clear that the process can be
repeated for the higher orders, but the two-loop case suffices
to explain the idea.
The one-loop corrections
to the insertion $\gh A_\Lambda$ \bref{fin anomaly 1} then suffice.
Higher order quantum dressings of this insertion
\bref{fin anomaly 1} are given, order by order in
perturbation theory, by  the well-known expression \ct{dewitt}
\be
   (\gh A_{\Lambda R}\cdot\Gamma_\Lambda)
   =\,:\exp\left\{\frac{i}{\hbar}\sum_{n=2}^{\infty}
   \frac{(-i\hbar)^n}{n!} G^{A_n\ldots A_1}_{\Lambda}
   \lder{}{\Phi^{A_1}}\ldots \lder{}{\Phi^{A_n}}\right\}: \gh A_{\Lambda R},
\label{q def}
\ee
with the colons ``$:\,\,:$'' indicating that functional derivates act only
on $\gh A_{\Lambda R}$ and where $G^{A_n\ldots A_1}_{\Lambda}$
stand for the connected correlation functions
$$
  G^{A_n\ldots A_1}_{\Lambda}=\restric{\left\{
   \lder{}{J_{A_n}}\ldots \lder{}{J_{A_1}}\left(
   \frac{i}{\hbar}\ln Z_{\Lambda}[J,\Phi^*]\right)\right\}}
   {J=J(\Phi,\Phi^*)},
$$
associated with our nonlocally regulated action $W_\Lambda$
\bref{w lambda}.
Up to two--loops the differential operator in \bref{q def} is given by
\be
  1+\frac{i\hbar}2 (S^{-1}_\Lambda)^{AB}
   \lder{}{\Phi^{B}}\lder{}{\Phi^{A}}+ O(\hbar^2)\equiv
   1+\hbar Q_1 + O(\hbar^2),
\label{q1 def}
\ee
where the ``complete'' propagator $(S^{-1}_\Lambda)^{AB}$ is the
inverse of the hessian of the classical part $S_\Lambda$ of $W_\Lambda$,
$(S_\Lambda)_{AB}= \frac{\partial_l\partial_r S_\Lambda}
{\partial\Phi^A\partial\Phi^B}$. The loop expansion of
the insertion $\gh A_\Lambda$ \bref{fin anomaly 1} up to second order
reads
$$
  \gh A_{\Lambda R}(\Phi,\Phi^*)=
  \left[\gh A_{1\Lambda}+ \hbar\gh A_{2\Lambda} + O(\hbar^2)\right]
  (\Phi+\bar\Psi_q,\Phi^*\gh\veps^2),
$$
with the functionals $\gh A_{1\Lambda}$, $\gh A_{2\Lambda}$
given by \ct{p95}
\bea
    \gh A_{1\Lambda}&=&
    \left[\Omega_S +i(M_1, S)\right],
\nonumber\\
    \gh A_{2\Lambda}&=&
    \left[\Omega_{M_1} +\frac{i}2 (M_1, M_1)+ i(M_2, S)\right],
\nonumber
\eea
in terms of the zero-th order $\Omega_S$ \bref{omega 0} and the
first order $\Omega_{M_1}$ in the loopwise expansion of
$\Omega_W$ \bref{final delta reg}%
\footnote{Note the presence of the cutoff in these expressions:
see footnote~\ref{Rnotation} on page~\pageref{Rnotation}.}.

Insertion of these expansions in \bref{q def} gives
the following form of the generating functional of the 1PI
diagrams with one insertion of $\gh A_\Lambda$:
\be
  (\gh A_{\Lambda R}\cdot\Gamma_\Lambda)=
  \gh A_{1\Lambda}(\Phi+\bar\Psi_q,\Phi^*\gh\veps^2)+
  \hbar\left[\gh A_{2\Lambda}(\Phi+\bar\Psi_q,\Phi^*\gh\veps^2)+
  Q_1 \gh A_{1\Lambda}(\Phi+\bar\Psi_q,\Phi^*\gh\veps^2)\right]+
  O(\hbar^2).
\label{2l a gamma}
\ee
Comparing this  with eq.\,\bref{claim}, the expression for the
one-loop anomaly provided by nonlocal regularization is immediatley
recognized:
\be
  \gh A_1= \lim_{\Lambda^2\rightarrow\infty}
  \gh A_{1\Lambda}(\Phi+\bar\Psi_q,\Phi^*\gh\veps^2)=
  \left\{\lim_{\Lambda^2\rightarrow\infty}
    \left[\Omega_S +i(M_1, S)\right]\right\}(\Phi,\Phi^*),
\label{oneloop anom bis}
\ee
and agrees with the form \bref{oneloop anom}  suggested by
the prescription \bref{fin anomaly 2}.

We now continue the investigation at two loop order.
There is of course the naively expected contribution
from the term $\gh A_{2\Lambda}$ in \bref{2l a gamma}.
This is associated with the contribution to the complete
two--loop anomaly due to the addition of counterterms $M_1$ and $M_2$.
In addition there is the one--loop correction to the one--loop anomaly of
the regularised theory $\gh A_{1\Lambda}$:
\bea
   Q_1 \gh
   A_{1\Lambda}(\Phi+\bar\Psi_q,\Phi^*\gh\veps^2)&=&
   \frac{i}2\left\{(\gh\veps^2)^A_{\,B} (S^{-1})^{BC}
   \left[(\gh A_{1\Lambda})_{CD}(\delta_\Lambda^{(0)})^D_{\,A}
     \phantom{\rder{\gh A_{1\Lambda}}{\Phi^D}
     \lder{(\delta_\Lambda^{(0)})^D_{\,A}}{\Phi^C}}
     \right.\right.
\label{1loop dressing}\\
    &+&   \left.\left.
   (-1)^{(D+1)C} \rder{\gh A_{1\Lambda}}{\Phi^D}
   \lder{(\delta_\Lambda^{(0)})^D_{\,A}}{\Phi^C}\right]\right\}
   (\Phi+\bar\Psi_q,\Phi^*\gh\veps^2)+O(\hbar)\, ,
\nonumber
\eea
where the $O(\hbar)$ results from replacing $\delta_\Lambda$ by
$\delta_\Lambda^{(0)}$ (see \bref{quantum delta} and \bref{delta lambda})
and use has been made of the classical analogs of \bref{relations der w}.

The limit $\Lambda^2\rightarrow\infty$
of the one--loop contribution in \bref{2l a gamma}  yields both
the one--loop dressings $(\gh A_1\cdot\Gamma_1)$
to the one-loop anomaly $\gh A_1$ \bref{oneloop anom}
{\em and} the genuine, local
two--loop anomaly $\gh A_2$, i.e.
$$
  \lim_{\Lambda^2\rightarrow\infty}
  \left[\gh A_{2\Lambda}(\Phi+\bar\Psi_q,\Phi^*\gh\veps^2)
  + (\gh A_{1\Lambda}(\Phi+\bar\Psi_q,\Phi^*\gh\veps^2)\cdot\Gamma_1)\right]=
  \gh A_2+(\gh A_1\cdot\Gamma_1).
$$
We repeat how the last term in this expression
is to be computed: first the one--loop
anomaly functional is obtained by the limiting process in the
previous step, see \bref{oneloop anom bis};
this expression is regularised ($\gh A_{1R}$), see \bref{nlchange};
this regularised expression is used in a regularised one--loop
computation (using $W_\Lambda$ and $\Gamma_\Lambda$);
finally,  the $\Lambda\rightarrow\infty$ limit is taken.
From \bref{q def} and \bref{q1 def}
we can compute this one--loop dressing as
$$
  (\gh A_1\cdot\Gamma_1) = \lim_{\Lambda^2\rightarrow\infty}
  \left[Q_1 \gh A_1(\Phi+\bar\Psi_q,\Phi^*\gh\veps^2)\right].
$$
This results in the following expression
for the genuine, local two--loop anomaly:
\be
   \gh A_2= \lim_{\Lambda^2\rightarrow\infty}
   \left[\gh A_{2\Lambda}(\Phi+\bar\Psi_q,\Phi^*\gh\veps^2)+
   Q_1 \left((\gh A_{1\Lambda}-\gh A_1)
   (\Phi+\bar\Psi_q,\Phi^*\gh\veps^2)\right)\right].
\label{2loop}
\ee
We recall that the  $\gh A_{2\Lambda}$ term is due to
the presence of counterterms. The remaining terms are
evanescent. Nevertheless, checking
the balance of inverse powers of $\Lambda$
(generically contained in evanescent functionals) with the positive powers
arising from the loop--divergence, one concludes that in some circumstances
they have a  finite limit.
This is, in this framework, the origin of the two--loop anomaly.

The finite limit of the  second term in \bref{2loop} is actually a
{\em local} functional. This corresponds to the locality of the
divergences of quantum field theory: it arises from such a divergence.
This fact allows, just as in the one--loop case, considerable
simplifications in the actual computations by taking into account
rather general features of the theory, like the form of the interaction
or dimensional analysis.
Therefore, equation \bref{2loop} is not only a perfectly legitimate
expression for the two--loop anomaly,
but can also be used to actually compute it.
Rather than elaborate on the circumstances that may give rise
to a non--zero result, we will illustrate the mechanism at work,
for chiral $W_3$ gravity, in section~\ref{w3example}.

\section{Higher Loop Consistency Conditions}
\label{hlcc}

\hspace{\parindent}%
The computation of one--loop anomalies is greatly facilitated by the fact
that they satisfy a consistency condition, the celebrated Wess--Zumino
condition \ct{wz}. In many cases this condition allows only a specific
functional
form as solution, thus reducing the actual computation of the anomaly to
the computation of a coefficient.
Having learned how to extract information on higher loop BRST
anomalies from the expression of the regulated BRST Ward identity
\bref{reg ward ident}, we now set out to derive
the corresponding consistency conditions for these higher loop anomalies.

In the present regularisation framework,
formal expressions become true equalities after
substituting for the local functionals their nonlocally
regulated counterparts. The regulated
version of the original FA consistency condition \bref{lcc},
$\sigma\gh A=0$, can thus be formulated by replacing $\gh A$ by its
regulated version $\gh A_{\Lambda R}$ \bref{fin anomaly 1},
i.e. $\sigma\gh A_{\Lambda R} =0$.
This is simply a consequence of the Jacobi identity for the regulated
effective action $\Gamma_\Lambda$
\be
    \left((\sigma\gh A_{\Lambda R})\cdot\Gamma_\Lambda\right)=
    \left((\gh A_{\Lambda R}\cdot\Gamma_\Lambda),\Gamma_\Lambda\right)=
    \frac{i}{2\hbar}
    \left((\Gamma_\Lambda,\Gamma_\Lambda),\Gamma_\Lambda\right)
     \equiv 0\, .
\label{jacobi reg}
\ee
Although this is an identity,
it also implies a restriction on the possible form
of the anomaly, as we now explain. Indeed, taking into account our previous
result \bref{claim2} and the finiteness of all the limits involved,
the following chain of equalities is then seen to hold
\bea
  \lim_{\Lambda^2\rightarrow\infty}
  \left((\sigma \gh A_R+ \gh A_R\gh A_{\Lambda R})
  \cdot\Gamma_{\Lambda}\right)&=&
  \lim_{\Lambda^2\rightarrow\infty}
  \left((\gh A_R\cdot\Gamma_\Lambda),\Gamma_\Lambda\right)=
\nonumber\\
  \left(\lim_{\Lambda^2\rightarrow\infty}
  (\gh A_R\cdot\Gamma_\Lambda),
  \lim_{\Lambda^2\rightarrow\infty}\Gamma_\Lambda\right)&=&
  \left(\lim_{\Lambda^2\rightarrow\infty}
  (\gh A_{\Lambda R}\cdot\Gamma_\Lambda),
  \lim_{\Lambda^2\rightarrow\infty}\Gamma_\Lambda\right)=
\nonumber\\
  \lim_{\Lambda^2\rightarrow\infty}
  \left((\gh A_{\Lambda R}\cdot\Gamma_\Lambda),\Gamma_\Lambda\right)=0.
\label{cchla}
\eea
However, whereas \bref{jacobi reg} is an identity that is valid even before
taking the $\Lambda\rightarrow\infty$ limit, this is no longer true for
\bref{cchla}. In particular it should be noted that the
limit of the insertion $(\sigma \gh A_R+ \gh A_R\gh A_{\Lambda R})$
by itself {\em does not vanish}! The reason is in fact the same as that
giving rise to the higher loop anomaly: one can not neglect the
effect of evanescent terms on loop diagrams,
since their quantum dressing may have a non--vanishing limit.
One should therefore keep equation \bref{cchla} as
it stands.

The content of \bref{cchla} can be disentangled using a loop-wise
expansion. To compute the antibracket in $\sigma\gh A_R$,
namely $(\gh A_R, W_\Lambda)$, it is expedient to  revert to
the variables $\Theta,\Theta^*$ in intermediate steps:
the canonical character of the transformation \bref{new variables}
simplifies the computation.
With the notation
$\tilde\gh A(\Phi,\Phi^*,\Psi,\Psi^*)= \gh A(\Theta,\Theta^*)$
one finds
(for $\tilde W$ see \bref{aux quantum action})
\bea
   (\bar\Psi^A_q, W_\Lambda)(\Phi,\Phi^*)&=&
   \restric{\lder{\tilde W}{\Psi^*_A}}{q}+
   \left[\gh K^{AB} \lder{}{\Phi^B}\frac12(W,W)\right]
  (\Phi+\bar\Psi_q,\Phi^*\gh\veps^2),
\nonumber\\
  (\gh A_{R}, W_{\Lambda})(\Phi,\Phi^*)&=&
  \restric{(\tilde \gh A, \tilde W)}{q}+
  \left[\rder{\gh A}{\Phi^A}\,
  \gh K^{AB}\,\lder{}{\Phi^B}\frac12(W,W)\right]
   (\Phi+\bar\Psi_q,\Phi^*\gh\veps^2),
\nonumber\\
  \restric{(\tilde\gh A, \tilde W)}{q}&=&
  \restric{(\gh A, W)(\Theta,\Theta^*)}{q}=
  (\gh A,W)(\Phi+\bar\Psi_q,\Phi^*\gh\veps^2).
\nonumber
\eea
The antibracket now becomes%
\footnote{One notes that the process of replacing functionals $F,G$
with their regulated version $F_R,G_R$ does not commute with taking
antibrackets: $(F_R,G_R)\neq(F,G)_R$. For the special case of $S_\Lambda$
however (i.e. $W_\Lambda$  without the counterterms),
one does obtain a simple formula, $(F_R,S_\Lambda)=(F,S)_R$, as can be
read off from \bref{nice anti}, since this equation is in fact valid
for arbitrary functionals $\gh A$.}
\be
   (\gh A_R, W_\Lambda)(\Phi,\Phi^*)= \left[(\gh A,W)+
   \rder{\gh A}{\Phi^A} \,\gh K^{AB}\, \lder{}{\Phi^B}\frac12(W,W)\right]
   (\Phi+\bar\Psi_q,\Phi^*\gh\veps^2).
\label{nice anti}
\ee

The explicit expression for $\Delta\gh A_R$, on the other hand,
computed along similar lines to that of $\Delta W_\Lambda$
in section~\ref{nrfa}, results in
\be
    \Delta \gh A_R(\Phi,\Phi^*)\equiv
    \left(\Omega_{\gh A}^{(i)}+ \Omega_{\gh A}^{(ii)}\right)
    (\Phi+\bar\Psi_q,\Phi^*\gh\veps^2),
\label{delta b}
\ee
with the functionals $\Omega_{\gh A}^{(i,ii)}$ given by
\bea
    &&\Omega_{\gh A}^{(i)}\equiv (-1)^{(A+1)}
  \left[\gh A^A_{\,B}(\delta_\Lambda)^B_{\,C}(\gh\veps^2)^C_{\,A}+
   W^A_{\,B} \left( \delta_\Lambda(\gh O \gh A) \delta_\Lambda \right)^B_{\,C}
  (\gh\veps^2)^C_{\,A}\right],
\label{omegab}\\
    &&\Omega_{\gh A}^{(ii)}\equiv
    \rder{\gh A}{\Phi^A} \, \gh K^{AB} \,
   \lder{\Omega_W}{\Phi^B},
\nonumber
\eea
where $(\gh O \gh A)^A_{\,B}= \gh O^{AC} \gh A_{CB}$
and for $\gh K$ see \bref{inverse k}.

Putting together \bref{delta b} and \bref{nice anti} one gets
$$
   (\sigma \gh A_R+\gh A_R\gh A_{\Lambda R})(\Phi,\Phi^*)=
   \left\{
   (\gh A, W)-i\hbar \Omega_{\gh A}^{(i)}
   -i\hbar\rder{\gh A}{\Phi^A} \, \gh K^{AB} \,
   \lder{\gh A_\Lambda}{\Phi^B} +\gh A\gh A_\Lambda\right\}
   (\Phi+\bar\Psi_q,\Phi^*\gh\veps^2).
$$
The antibracket and the $\Omega_{\gh A}^{(i)}$ term in this expression
may then be seen as the regularised version of the naively expected terms.
The other terms
\be
  \left(\gh A\,\gh A_\Lambda
   -i\hbar\rder{\gh A}{\Phi^A} \, \gh K^{AB} \,
   \lder{\gh A_\Lambda}{\Phi^B}\right)
   (\Phi+\bar\Psi_q,\Phi^*\gh\veps^2),
\label{extra contrib}
\ee
are absent from the naive expressions, for the simple
reason that they would vanish identically because of fermi statistics if
we would not distinguish between $\gh A$ and $\gh A_\Lambda$.
Their contribution to the consistency
condition is in fact of order $\hbar^2$, and therefore only comes into
play when considering the condition for $\gh A_3$ or higher.
The reason is that the product term does not generate
tree level or $O(\hbar)$ contributions to the 1PI
functional, whereas the second term, with an explicit factor $\hbar$, is
evanescent and only contributes through quantum dressings.
We will limit ourselves in the rest of this
section to the first nontrivial consistency condition, that for the
two-loop anomaly, so we do not consider \bref{extra contrib} further.
The power series expansion can  then be carried out in much the same way
as  in the previous section. The result is
\bea
  &\displaystyle{
  \left((\gh A, W)-i\hbar \Omega_{\gh A}^{(i)}\right)
   (\Phi+\bar\Psi_q,\Phi^*\gh\veps^2)= }&
\nonumber\\
  &\displaystyle{
   \left\{(\gh A_1, S)+
   \hbar\left[(\gh A_2, S)+ (\gh A_1, M_1) -i \Omega_{\gh A_1}\right]
   +O(\hbar^2) \right\}
   (\Phi+\bar\Psi_q,\Phi^*\gh\veps^2).}&
\label{1loop insert}
\eea
Here $\Omega_{\gh A_1}$ is the lowest order term
in the loop-wise expansion of $\Omega_{\gh A}^{(i)}$
\bref{omegab}
\be
    \Omega_{\gh A_1}= (-1)^{(A+1)}
  \left[(\gh A_1)^A_{\,B}(\delta_\Lambda^{(0)})^B_{\,C}(\gh\veps^2)^C_{\,A}+
   S^A_{\,B} \left( \delta_\Lambda^{(0)}(\gh O \gh A_1)
   \delta_\Lambda^{(0)} \right)^B_{\,C} (\gh\veps^2)^C_{\,A}\right],
\label{omegaa1}
\ee
the notational convention of footnote~\ref{notation} has been used,
and $\delta_\Lambda^{(0)}$ is given by \bref{delta lambda}.
The lowest order term in this expression should vanish
according to condition \bref{cchla}
$$
(\gh A_1, S)=0,
$$
which is the usual Wess--Zumino (one--loop) consistency condition.
This same equation plays an important role in simplifying
the two--loop anomaly condition, since it
implies that when inserting \bref{1loop insert} in $\Gamma_\Lambda$,
equation \bref{cchla}, no quantum corrections have to be considered.
As a consequence, at this order, the potentially nonlocal \bref{cchla}
turns into a {\em local} equation, which is
the final expression for the consistency condition on the genuine
two--loop anomaly:
\be
  \left\{(\gh A_2, S)+
  \lim_{\Lambda^2\rightarrow\infty}
  \left[(\gh A_1, M_1) -i \Omega_{\gh A_1}\right]\right\}
  (\Phi,\Phi^*)=0.
\label{cchla2}
\ee
The remaining limit is to be understood as before in the one--loop case:
the two terms separately can diverge as $\Lambda\rightarrow\infty$. There
is a difference with \bref{oneloop anom} however.
In that case, if  a divergence is present, it
imposes that $M_1$ be chosen such that the cancellation occurs and the
difference remains finite. Here, $M_1$ has already been fixed,
precisely by that
one--loop criterion, so the cancellation in \bref{cchla2} should be
automatic. Although ultimately this is implied by the finiteness of
$\Gamma$ if the theory is renormalisable, it would be nice to have a more
direct proof of this fact from the explicit expressions
\bref{omegaa1} and \bref{oneloop anom}.

Summarizing, we have verified that in this nonlocally regulated
framework not only the one--loop but also
the two--loop anomalies are constrained by a {\it local}
consistency condition, \bref{cchla2}, which is very similar in form to
that obtained in the unregulated formulation, \bref{twoloop}.

Consistency conditions for higher loop anomalies can now be obtained as
well by  repeating the procedure demonstrated above.
Although the analysis of their explicit form
would  require a  considerable amount of work,
it may be inferred from  \bref{cchla}
that they should also share with \bref{cchla2} the generic local form
\be
     (\gh A_p, S)-
     \gh U_p(\gh A_1,\ldots,\gh A_{p-1};S, M_1,\ldots,M_{p-1})=0,
     \quad\quad p\geq 2.
\label{cchlap}
\ee
Potential nonlocalities arising in the process should again vanish as a
consequence of lower order consistency conditions.
Equation \bref{cchlap} differs from the standard
Wess-Zumino consistency condition  by the presence of extra
inhomogenities $\gh U_p$.
The cohomological characterization of higher loop parts of the
BRST anomaly by means of eqs.\,\bref{cchlap} is therefore different
from the lowest order anomaly. The solutions of the Wess--Zumino
condition corresponds to classifying local cohomology
classes of the nilpotent operator $\hat\delta=(\cdot, S)$
\bref{gen brst} at ghost number one.
At higher order this same classification will also be involved in
characterising a possible arbitrariness in the solution of \bref{cchlap},
but also a particular solution must be found.
The algebraic analysis of the higher loop relations
\bref{cchlap} and the characterization of higher loop BRST anomalies
falls however outside the scope of this paper.

\section{The Two--loop Anomaly in Chiral $W_3$ gravity}
\label{w3example}

\hspace{\parindent}%
In the previous sections we have deduced the form of genuine,
local higher loop anomalies and of their consistency conditions
from the quantum master equation \bref{anomaly}.
Now we proceed to an illustration, and an explicit verification
of the resulting expressions \bref{2loop} and \bref{cchla2},
in the example of chiral $W_3$ gravity --for
which the corresponding BRST anomaly gets contributions up to two loops.

\subsection{Nonlocally Regulated Chiral $W_3$ Gravity}
\label{NLRCW3}

\hspace{\parindent}%
We use the conventions and notations of refs.\,\cite{vp94} and \ct{p95}.
This subsection contains selected material from these papers,
to which we refer the reader  for the actual
construction of the proper solution of the master equation
and the nonlocalization process for $W_3$ gravity respectively.

Chiral $W_3$ gravity \cite{hull90a} can be realised as
a system of $D$ two dimensional
scalar fields $\phi^i$, $i=1,\ldots, D$, coupled to
gauge fields $h$ and $B$ through the chiral spin-2 and spin-3 currents
$$
  T=\frac12(\partial\phi^i)(\partial\phi^i),\quad\quad
  W=\frac13 d_{ijk} (\partial\phi^i) (\partial\phi^j) (\partial\phi^k),
$$
defined in terms of the usual combinations of space-time derivatives
$\partial=\partial_+$, $\bar\partial=\partial_-$, with
$x^{\pm}=\frac1{\sqrt2}(x^1\pm x^0)$, and of a constant, totally symmetric
tensor $d_{ijk}$ satisfying the identity
$$
   d_{i(jk}d_{l)mi}=k\delta_{(jl}\delta_{k)m},
$$
for an arbitrary, but fixed parameter $k$.

The proper solution of the classical master equation for this system
\cite{vp94} is
\bea
   S=\int\dif^2 x\hspace{-5mm}&&\left\{
   \left[-\frac12(\partial\phi^i)(\bar\partial\phi^i)
   +b(\bar\partial c)+ v(\bar\partial u)\right]\right.
\nonumber\\
   &&+\phi^*_i\left[c(\partial\phi^i)+
   u d_{ijk} (\partial\phi^j) (\partial\phi^k)
   -2 k b (\partial u) u (\partial\phi^i)\right]
\nonumber\\
   &&+b^*\left[-T +2 b(\partial c)+ (\partial b) c
   + 3 v(\partial u) + 2(\partial v) u \right]
\nonumber\\
   && +v^*\left[-W +2 k T b(\partial u) +2 k \partial(T b u)
   + 3 v(\partial c) + (\partial v) c \right]
\nonumber\\
   && \left.
   +c^*\left[ (\partial c) c
   +2 k T (\partial u) u\right]
   +u^*\left[2(\partial c) u- c(\partial u)\right]\right\}
\nonumber\\
   &=& \gh S(\Phi) +\Phi^*_A R^A(\Phi),
\label{w3gfa}
\eea
where $\{c, u\}$ stand for the ghosts corresponding to spin-2 and spin-3
gauge symmetries; $\{b,v\}$, for their associated antighosts; and
$\{\phi^*_i, c^*, u^* ,b^*, v^*\}$, for the corresponding antifields. This
gauge--fixed action is obtained by performing a canonical transformation
from the classical basis of fields and antifields
$\{\phi^i, h, B, c, u;\phi^*_i, h^*, B^*, c^*, u^*\}$
to the so-called gauge--fixed basis:
$$
   \{h, h^*, B, B^*\}\rightarrow
   \{b=h^*, b^*=-h, v=B^*, v^*=-B\}.
$$
In this way, the classical interaction is completely
contained in the antifield dependent part, $\Phi^*_A R^A(\Phi)$, so that
antifields play the role of coupling constants.

Nonlocal regularization of the quantum theory stemming from the proper
solution \bref{w3gfa} is described in \ct{p95}. The propagating fields are
ordered in a vector $\Phi^A=\{\phi^i; b, v; c, u\}$.
The kinetic operator $F_{AB}$ in \bref{original action} is
$$
 F_{AB}=
 \left(\begin{array}{ccc}
    \partial\bar\partial\,\delta_{ij} & 0 & 0 \\
    0 & 0 & \unity\,\bar\partial \\
    0 &  \unity\,\bar\partial  &  0
  \end{array}\right),
$$
where $\unity$ stands for the identity in the spin 2 (spin 3) ghost
sector. The operator $(T^{-1})^{AB}$ is chosen as
$$
 (T^{-1})^{AB}=
 \left(\begin{array}{ccc}
    \delta^{ij} & 0 & 0 \\
    0 & 0 & \unity\,\partial\\
    0 & \unity\,\partial  & 0
  \end{array}\right),
$$
and yields a simple regulator $\gh R^A_B$ \bref{regulator} and
smearing operator $\gh\veps^A_{\,B}$ \bref{smearing op}:
$$
   \gh R^A_{\,B}= \partial\bar\partial\,\delta^A_{\,B},\quad\quad
   \gh\veps^A_{\,B}=
   \exp\left(\frac{\partial\bar\partial}{2\Lambda^2}\right)
   \delta^A_{\,B}\equiv\gh\veps \,\delta^A_{\,B}.
$$
Finally, the shadow propagator $\gh O^{AB}$ \bref{shadow propagator} is
\be
   \gh O^{AB}=
   \left(\begin{array}{ccc}
   \gh O & 0 & 0 \\
   0 & 0 & \unity\gh O \partial\\
   0 & \unity \gh O \partial  & 0
   \end{array}\right),
   \quad\quad\mbox{with} \quad\quad
   \gh O\equiv \frac{(\gh\veps^2-1)}{\partial\bar\partial}=
   \int^1_0\frac{\dif t}{\Lambda^2}\,
   \exp\left(t\frac{\partial\bar\partial}{\Lambda^2}\right).
\label{ShadowPropW3}
\ee

To complete the specification of the regulated theory we have to provide
the counterterms $M_p$. These are generically necessary to ensure
finiteness of the theory. However, chiral $W_3$ gravity is ''finite'', as
is well known, so that here inclusion of such counterterms
is in principle not necessary. One might still want to include finite
counterterms  (to preserve BRST invariance as far as possible),
but we will not do this here.
Then $W$ is just the same as the  classical action $S$
\bref{w3gfa}, and algebraic computations involving the operator $\Delta$
simplify accordingly. In the present case, we have for $S^A_{\,B}$
\be
  S^A_{\,B}= \left(\begin{array}{ccccc}
    c^i_j \partial & -2k(\partial u) u (\partial\phi^i) & 0 &
    (\partial\phi^i) & u^i\\
    -(\partial\phi_j)\partial & -(c\partial)_2 & -2(u\partial)_{3/2} &
    (b\partial)_1 & 3(v\partial)_{1/3}  \\
    -u_j\partial & -2k[ T (u\partial)_2+u(\partial T)] & -(c\partial)_3 &
    3(v\partial)_{1/3} &  4 k (bT\partial)_{1/2} \\
    2k(\partial u) u (\partial\phi_j) \partial &  0 & 0 &
    -(c\partial)_{-1} & -2 k T (u\partial)_{-1}\\
    0 & 0 & 0 & -2 (u\partial)_{-1/2} & -(c\partial)_{-2}
   \end{array}\right),
\label{wsab}
\ee
with $c^i_j$ and $u^i$ the operators
\bea
     c^i_j&=& \left[c\delta^i_j -2k b(\partial u) u\delta^i_j+
     2 u d^i_{\,jk}(\partial\phi^k)\right],
\label{cij}\\
     u^i&=& d^i_{\,jk} (\partial\phi^j) (\partial\phi^k)
     -2 k \left[b (\partial u) (\partial\phi^i)
                +(b (\partial\phi^i) u\partial)_1\right],
\nonumber
\eea
and where $(F(\Phi,\Phi^*)\,\partial)_n$ stands for the shorthand notation
$$
   (F\,\partial)_n= F\,\partial+ n (\partial F),
   \quad\quad (F\,\partial)_n^\dagger=
   -[F\,\partial+ (1-n) (\partial F)]=-(F\,\partial)_{1-n}.
$$
In much the same way, the operator $(I_{\rm cl})_{AB}$ reads
\be
   (I_{\rm cl})_{AB} = \left(\begin{array}{ccccc}
  \partial\,h^*_{ij}\,\partial & (g^*_i\partial)_1  & 0 &
  -(\phi^*_i\partial)_1 & (q^*_i\partial)_1\\
  g^*_j\,\partial & 0 & 0 & (b^*\partial)_{-1} & r^*\\
  0 & 0 & 0 & 2(v^*\partial)_{-1/2} & (b^*\partial)_{-2}\\
  -\phi^*_j\,\partial & (b^*\partial)_2 & 2 (v^*\partial)_{3/2} &
  2(c^*\partial)_{1/2} & -3(u^*\partial)_{2/3}\\
  q^*_j \, \partial & - (r^*)^\dagger & (b^*\partial)_{3} &
  -3(u^*\partial)_{1/3} & 2(p^*\partial)_{1/2}
   \end{array}\right).
\label{w tilde iab}
\ee
The following abbreviations have been used for the matrix elements,
all linear quantities in the antifields:
\bea
   h^*_{ij}&=&\delta_{ij}
   \left[ b^* + 2kb(u(\partial v^*)-v^*(\partial u))
   +2k c^*u(\partial u)\right]
   -2 d_{ij}^{\,\,\,\,k}
   \phi^*_k u +2 v^* d_{ijk} (\partial\phi^k),
\label{hij}\\
    g^*_i&=& 2k\left[\phi^*_i(\partial u) u+
   (v^*(\partial u) -u(\partial v^*))(\partial\phi_i)\right],
\nonumber\\
    q^*_i&=&-2 \phi^*_j d^j_{\,ik}(\partial\phi^k)+
    2k\left[\left( \partial(v^*b (\partial\phi_i)+u\phi^*_i b)\right)+
    b(\partial\phi_j)(v^*\partial)_1+\phi^*_i b (u\partial)_1\right],
\nonumber\\
    r^*&=& 2k\left[T(v^*\partial)_{-1}
    -\phi^*_i(\partial\phi^i)(u\partial)_{-1}\right],
\nonumber\\
    p^*&=& 2k\left[T c^*-\phi^*_i(\partial\phi^i)b\right].
\nonumber
\eea

In the absence of counterterms, the nonlocally regulated
anomaly $\gh A_{\Lambda R}$ \bref{fin anomaly 1} simplifies to
the regulated trace \bref{omega 0}:
\be
  \gh A_{\Lambda R}(\Phi,\Phi^*)= \Delta S_\Lambda(\Phi,\Phi^*)=
  \Omega_S(\Phi+\bar\Psi_0,\Phi^*\gh\veps^2).
\label{delta s w3}
\ee
According to eq.\,\bref{claim},
all the information about the anomalies in this theory is
contained in this quantity and its quantum corrections:
\bea
  \lim_{\Lambda^2\rightarrow\infty}(\Delta S_\Lambda\cdot\Gamma_\Lambda)&=&
  \lim_{\Lambda^2\rightarrow\infty} \left[(\Delta S_\Lambda)
  +\hbar (Q_1 \Delta S_\Lambda) + O(\hbar^2)\right]
\nonumber\\
  &=&\gh A_1+ \hbar\left[ \gh A_2+
  (\gh A_1\cdot\Gamma_1)\right] + O(\hbar^2).
\label{anom w3}
\eea

We now disentangle from this expression the one-loop anomaly
(a computation performed in \ct{p95}), and then
the two--loop anomaly.

\subsection{One--loop anomaly}

\hspace{\parindent}%
The relevant local contributions in the loop expansion \bref{anom w3}
can be inferred from considerations of dimensions and spin: they have
the general form
$$
  \Lambda^{-2(m-n)} (\Phi^*)^{m+1} \bar\partial^{-n}
   F_{m,n}(\Phi;\partial),\quad\quad
    m, n=0, 1, \ldots,\quad\quad m\geq n.
$$
By locality, $n=0$ for the $\gh A_i$ terms: intrinsic
nonlocalities of the form $(\bar\partial)^{-n}$, $n>0$ are
associated with quantum dressings of such local anomalies.
Also, in the limit $\Lambda\rightarrow\infty$ only
terms with $m=0$ remain, which are therefore linear in the antifields.
Evaluating these linear terms proceeds by expansion of $\Omega_S$
\bref{delta s w3},
given in terms of $S^A_{\,B}$, \bref{wsab}, and $(I_{\rm cl})_{AB}$,
\bref{w tilde iab}:
\be
   \Omega_S^{(1)}=
  \left[ \gh\veps^2\, S^A_{\,B}\, \gh O^{BC}\,(I_{\rm cl})_{CA}\right],
  \qquad\qquad
  \Omega_S= \Omega_S^{(1)}+ O((\Phi^*)^2) \,.
\label{gen omega0}
\ee
Separating the terms corresponding to the different antifields,
upon taking the limit $\Lambda^2\rightarrow\infty$, five
different contributions are obtained for the final expression for the
one--loop anomaly:
\be
  \gh A_1=
  \lim_{\Lambda^2\rightarrow\infty}\Omega_S^{(1)}=
  \gh A_1^{(i)}+ \gh A_1^{(ii)}+ \gh A_1^{(iii)}+
  \gh A_1^{(iv)}+ \gh A_1^{(v)}.
\label{w3 1loop}
\ee
The different contributions are given by \ct{p95}
\bea
  \gh A_1^{(i)} &=&\frac{i}{24\pi} \int\dif^2 x\,
   c^{ij}\,\partial^3\, h^*_{ij},
\nonumber\\
  \gh A_1^{(ii)} &=&
 \frac{-100\,i}{24\pi} \int\dif^2 x\, c\,\partial^3\, b^*,
\nonumber\\
  \gh A_1^{(iii)} &=& \frac{ik}{2\pi} \int\dif^2 x\,
   (v^*(\partial u) -u(\partial v^*))(\partial\phi^i)(\partial^3\phi^i),
\label{A1-5}\\
  \gh A_1^{(iv)} &=&\frac{ik}{6\pi} \int\dif^2 x\,T\left[
    5 (\partial^3 u) v^* -12 (\partial^2 u)(\partial v^*)
    +12 (\partial u)(\partial^2 v^*) -5 u(\partial^3 v^*)\right],
\nonumber\\
  \gh A_1^{(v)} &=&
    \frac{-ik}{6\pi} \int\dif^2 x\,\phi^*_i\left[
     6\partial\left(u (\partial u) (\partial^2\phi^i)\right)
     +9 (\partial^2 u) (\partial u) (\partial\phi^i)
     +8  u (\partial^3 u) (\partial\phi^i) \right],
\nonumber
\eea
and are in complete agreement with previous computations in the
literature, using either PV regularization \ct{vp94} or standard
conformal field theory techniques \cite{hull91}.

For the computation of the two--loop anomaly however, the limiting
expression for the one--loop anomaly is not sufficient: as emphasized in
the general section, we need information that is only contained
in the expression for the regulated anomaly {\em before} the limit.
The corresponding expressions for the different parts in $\Omega_S^{(1)}$,
$$
   \Omega_S^{(1)}= \Omega_S^{(1,i)}+ \Omega_S^{(1,ii)}+
   \Omega_S^{(1,iii)}+ \Omega_S^{(1,iv)}+ \Omega_S^{(1,v)},
$$
were also obtained in the course of the computations of \ct{p95}, but not
recorded there, so we give them now:
\bea
   \Omega_S^{(1,i)} &=&\frac{-i}{2\pi} \int\dif^2 x\,
   c^{ij}\, \bar\gh O(1,1;\Lambda^2)\, \partial^3\, h^*_{ij},
\label{Omega1-5}\\
   \Omega_S^{(1,ii)} &=& \frac{i}{2\pi}
   \int\dif^2 x\, c\,
    [4\bar\gh O(1,1;\Lambda^2)-16\bar\gh O(0,0;\Lambda^2)]
    \, \partial^3\, b^*,
\nonumber\\
   \Omega_S^{(1,iii)} &=& \frac{ik}{\pi} \int\dif^2 x\,
   (\partial^3\phi^i)\bar\gh O(0,0,\Lambda^2)\,
   \left[(v^*(\partial u) -u(\partial v^*))(\partial\phi^i)\right],
\nonumber\\
   \Omega_S^{(1,iv)}&=&\frac{-ik}{\pi} \int\dif^2 x\,
    T\,\left\{ v^*
    \left[4\bar\gh O(1,1;\Lambda^2)-\bar\gh O(0,0;\Lambda^2)\right]
    (\partial^3 u)\right.
\nonumber\\
    &&\left.\hspace{25mm}
    +4 (\partial v^*)\bar\gh O(0,0;\Lambda^2)(\partial^2 u)
    -(u\leftrightarrow v^*) \right\},
\nonumber\\
   \Omega_S^{(1,v)} &=&
    \frac{ik}{\pi} \int\dif^2 x\,\phi^*_i\left\{
     (\partial u)\, u\,\bar\gh O(0,0;\Lambda^2)\, (\partial^3\phi^i)+
     \bar\gh O(0,0;\Lambda^2)\, \partial^2
     \left( (\partial u) u (\partial\phi^i)\right)\right.
\nonumber\\
     &&\left.+4 (\partial\phi^i)(\partial u)
     \bar\gh O(0,0;\Lambda^2) (\partial^2 u)-
     4 (\partial\phi^i) u
     \left[\bar\gh O(0,0;\Lambda^2)+ \bar\gh O(0,1;\Lambda^2)\right]
     (\partial^3 u)\right\},
\nonumber
\eea
where again the abbreviations $c^i_j$ \bref{cij}, $h^*_{ij}$ \bref{hij}
were used, and the operator $\bar\gh O(n,m;\Lambda^2)$ is defined as
\be
   \bar\gh O(n,m;\Lambda^2)=
   \int_0^1\dif t\,\frac{(-1)^m\,t^m}{(1+t)^{n+m+2}}
   \exp\left(\frac{t}{1+t}\,\frac{\partial\bar\partial}{\Lambda^2}\right).
\label{o operator}
\ee
In the limit, the operators $\bar\gh O(n,m;\Lambda^2)$
in these expressions become numerical coefficients, resulting in
\bref{A1-5}.

\subsection{Two--loop anomaly}

\hspace{\parindent}%
The evaluation of the two--loop anomaly is greatly facilitated by the fact
that, as noted in the previous subsection, it depends linearly on the
antifields. Since $\Omega_S^{(1)}$ is already linear,
see \bref{gen omega0},
one may neglect the antifield dependence in the rest of
\bref{1loop dressing}
$$
  (S^{-1})^{AB}= (F^{-1})^{AB}+ O(\Phi^*),\quad\quad
  (\delta_\Lambda^{(0)})^A_{\,B}= \delta^A_{\,B}+ O(\Phi^*).
$$
The local part that eventually may arise from
the one--loop dressing $(Q_1 \Delta S_\Lambda)$ of the insertion
\bref{delta s w3} is then
$$
  Q_1\,\Delta S_\Lambda= \frac{i}{2}
  \left[\gh\veps^2 (F^{-1})^{AB}
  \left(\Omega_S^{(1)}\right)_{BA}\right]
   (\Phi+\bar\Psi_0,\Phi^*\gh\veps^2)+ O((\Phi^*)^2).
$$
We require the full $\Lambda$ dependence of $\Omega_S^{(1)}$
\bref{gen omega0},
which was given in \bref{Omega1-5}.
Similar arguments determine the possible local contributions
generated by the genuine, local one-loop anomaly $\gh A_1$
\bref{w3 1loop} to be of an analogous form
$$
  Q_1 \gh A_1(\Phi+\bar\Psi_0,\Phi^*\gh\veps^2)= \frac{i}{2}
  \left[\gh\veps^2 (F^{-1})^{AB}(\gh A_1)_{BA}\right]
   (\Phi+\bar\Psi_0,\Phi^*\gh\veps^2)+ O((\Phi^*)^2).
$$
The two--loop anomaly resulting from this analysis of \bref{2loop},
and in the absence of counterterms, is given by
\be
   \gh A_2= \lim_{\Lambda^2\rightarrow\infty} \left\{\frac{i}{2}
  \left[\gh\veps^2 (F^{-1})^{AB}
  \left(\Omega_S^{(1)}-\gh A_1\right)_{BA} \right]\right\}.
\label{2loopw3}
\ee
The interpretation of this formula is that the two--loop anomaly results
by contracting external lines (the factor $F^{-1}$)
in a regularised fashion (the factor $\gh\veps^2$).
The expression that undergoes this contraction is however
not the one--loop anomaly (which is $\gh A_1$), but rather the
(evanescent) difference between the {\em regulated} one-loop anomaly
and the one--loop anomaly itself.

The two loop anomaly can be computed from \bref{2loopw3} in a
straightforward manner. We will, for illustrative purposes, pick out a
specific term and show in detail the mechanism to generate $\gh A_2$.
Afterwards, we will give the formulas needed for the
complete calculation, with considerably less detail.

\begin{figure}
\epsfxsize=140mm
\epsffile[54 701 424 783]{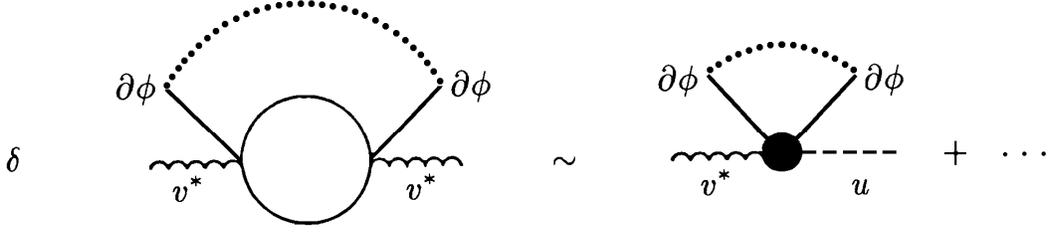}
\caption{The diagram on the left (without the dotted line) represents a
contribution to $\Gamma_1$ with two external matter lines. Its variation
$\delta$ yields (a.o.) the one--loop anomaly represented by the diagram
on the
right (again without the dotted line). Contracting the external matter
lines, the diagrams contribute to $\Gamma_2$ and the two--loop
anomaly respectively. In that case it is essential that the vertex on
the r.h.s. is nonlocalised:
the fat blob stands for the smearing operator
$\bar\gh O$ of eq.\,\bref{o operator}. }
\end{figure}

A specific non--zero contribution to the two--loop
anomaly arises from  the last term of $h^*_{ij}$ in \bref{hij}
via $\Omega_S^{(1,i)}$ of \bref{Omega1-5} with the last term in $c_{ij}$,
see \bref{cij}.
The corresponding diagrams are given in figure~1. At one
loop, in $\Gamma$, the  diagram with two external $\phi$ lines
and two factors $v^*$ gives rise to an anomalous variation,
in $(\Gamma,\Gamma)$, which contains
(among others) the term under consideration.
At the two--loop level,  a contribution
arising from ''closing the $\phi$ propagator'' can similarly give rise
to a two loop anomaly. In the figure, the smeared out vertex symbolizes
the fact that not just the local $\Lambda\rightarrow\infty$ part
is to be taken (in which case the blob would be a point),
but one should keep non--leading parts also.
The corresponding double momentum integral for the exact expression, in
non--local regularisation, takes the form
$$
   \int \frac{\dif^2 k}{(2\pi)^2}\,\int_0^1 \dif t\,
    \frac{t}{(t+1)^4}\, \frac{(p+k)^3\,k^2}{k\,\bar k} \,
    \exp\left(-\frac{t}{t+1}\frac{|p+k|^2}{2\Lambda^2}\right)
    \exp\left(-\frac{|k|^2}{2\Lambda^2}\right).
$$
One recognises the three momentum factors corresponding to the
three derivatives on $h^*_{ij}$ in $\Omega_S^{(1,i)}$, the two momentum
factors on the contracted $\phi$-line,
and the scalar field propagator. There are two cutoff factors,
one corresponding to the explicit $\veps^2$ in formula \bref{2loopw3}
and the other to the $\Lambda$ dependence of
$\Omega_S^{(1,i)}$ as given by the factor $\bar\gh O(1,1;\Lambda^2)$.
The second term in \bref{2loopw3}, corresponding to removing the one--loop
dressing of the one--loop anomaly, has exactly the same form but without
this last factor, i.e.
$\bar\gh O(1,1;\Lambda^2)\rightarrow \bar\gh O(1,1;\infty)=-1/12$.
The integral is well--defined in both cases.
In the latter case, the angular $k$-integral shows that it vanishes.
In the former case, the integral is equal to
\be
   \int \frac{\dif^2 k}{(2\pi)^2}\,\int_0^1 \dif t\,
   \frac{t}{(t+1)^4}\, \frac{(p+k)^3\,k^2}{k\,\bar k} \,
   \exp\left(-\frac{2t+1}{t+1}\frac{|k|^2}{2\Lambda^2}\right)
   \exp\left(-\frac{t}{t+1}
   \frac{|p|^2+p\bar k+\bar p k}{2\Lambda^2}\right).
\label{to-expand}
\ee
A variety of methods exist to compute
the $\Lambda\rightarrow\infty$ limit of this integral.
It is instructive to evaluate it by a power series expansion of
the last factor in \bref{to-expand}.
The first few terms are again zero due to the angular integration.
In fact, only terms with $\bar k^n$ with $n=2,3,4,5,$ give nonvanishing
contributions. The  integral then becomes, in the limit,
equal to a numerical factor times $p^5$.

Several aspects of this sample calculation merit some comment. First, it
is seen quite explicitly that taking the $\Lambda\rightarrow\infty$ limit
too early, i.e.\,replacing $\bar\gh O(1,1;\Lambda^2)$ in $\Omega_S^{(1,i)}$
by its limiting value, would be a mistake and give a vanishing overall
result. 
From the reasoning above it follows that dressing of terms
with inverse powers of $\Lambda^2$ result in finite contributions.
This is the evanescent aspect of the difference.
A second comment is that the
intuitive idea, that the two--loop anomaly results from a $\phi$-dependent
one--loop anomaly by closing the $\phi$--loop, has some truth but is
certainly not the complete truth.
Indeed, closing this loop on the final one--loop anomaly
in fact gives a {\em vanishing result} (recall the discussion of the second
term of \bref{2loopw3} in the instance above). It does
arise from such a diagram however if one takes the one--loop anomaly in
its regulated form {\em before taking the limit}: the non--zero
contribution is due to terms in the one--loop anomaly that have inverse
powers of the cutoff $\Lambda$.
A third comment is that the power series expansion as performed on the
integral above automatically yields a polynomial in the momentum $p$.
This attests the fact that the resulting expression is local.

Having illustrated the mechanism by considering one term in detail,
we now give a more complete account.
Starting again from \bref{2loopw3},
due to the simple form of the free propagator $(F^{-1})^{AB}$
only a few of the matrix elements of $\Omega_S^{(1)}$ are involved,
namely
\bea
   \left(\Omega_S^{(1)}\right)_{ij}&=&
   \frac{2i}{\pi}d_{ij}
   \left[\partial\,u\,\bar\gh O(1,1;\Lambda^2)\partial^3\,v^*\,\partial-
   \partial\,v^* \bar\gh O(1,1;\Lambda^2) \partial^3\, u\, \partial\right]
\nonumber\\
    &&+\frac{ik}{\pi}\delta_{ij}
    \left[\bar\gh O(0,0;\Lambda^2)
    \partial^3 (u(\partial v^*) - v^*(\partial u)) \partial
    +\partial (u(\partial v^*) - v^*(\partial u)) \partial^3
    \bar\gh O(0,0;\Lambda^2) \right]
\nonumber\\
    &&+\frac{ik}{\pi}\delta_{ij}\left\{
  \partial\left[v^*
  \left(4\bar\gh O(1,1;\Lambda^2)-\bar\gh O(0,0;\Lambda^2)\right)
  (\partial^3u)\right]\partial\right.
\nonumber\\
   &&\left. \hspace{13mm} +4 \partial \left[
   (\partial v^*)(\bar\gh O(0,0;\Lambda^2)\partial^2 u)\partial\right]
   -(u\leftrightarrow v^*)\right\},
\nonumber\\
   \left(\Omega_S^{(1)}\right)_{b c}&=& \frac{iDk}{\pi}
    (u(\partial v^*) - v^*(\partial u)) \bar\gh O(1,1;\Lambda^2)
    \partial^3, \quad\quad
   \left(\Omega_S^{(1)}\right)_{c b}=
   -\left(\Omega_S^{(1)}\right)^\dagger_{b c},
\nonumber
\eea
(with $d_{ij}\equiv d_{ilm} d_j^{\,\,lm}$)
and their local counterparts $\left(\gh A_1\right)_{AB}$  obtained
by  taking the limit $\Lambda^2\rightarrow\infty$ (some of these
are given later, in eq.\,\bref{aij}).
The expression \bref{2loopw3} reduces to the functional trace
\be
   \gh A_2(u,v^*)= \lim_{\Lambda^2\rightarrow\infty}
   \left\{ \frac{i}{2} \tr\left[\gh\veps^2
   \left(\frac{\delta^{ij}}{\partial\bar\partial}
   \left(\Omega_S^{(1)}-\gh A_1\right)_{ji}+
   \frac1{\bar\partial} \left(\Omega_S^{(1)}-\gh A_1\right)_{bc}+
   \left(\Omega_S^{(1)}-\gh A_1\right)_{bc} \frac1{\bar\partial}
   \right)\right] \right\},
\label{2loopw3bis}
\ee
in which only a functional dependence on the fields $u$ and $v^*$ appears.

Two types of terms are present,
proportional to one of the two parameter combinations
$d^2= d_{ijk} d^{ijk}$ and $kD$ that arise upon taking traces with respect
the discrete matter indices. Terms proportional to $kD$ are
always of the form
$$
   \tr\left[F(x)\bar\gh O(n,m;\Lambda^2)\frac{\partial^3}{\bar\partial}
   \gh\veps^2\right] \quad \mbox{or} \quad
   \tr\left[G(x)\frac{\partial^3}{\bar\partial} \gh\veps^2\right],
$$
where  $F(x)$ and  $G(x)$ are  {\it functions} of the fields $u$ and
$v^*$. The corresponding traces vanish, as can be seen by writing them
out in momentum space and doing the angular integration, along the same
lines as in the sample calculation above.
The same result holds true for all the terms (including the
$d^2$  contributions) generated by the one--loop anomaly $\gh A_1$.
A nonvanishing result
for the traces we are dealing with may only result by the
presence of  $\bar\gh O(n,m;\Lambda^2)$ operators \bref{o operator}
{\it between} the fields $u$ and $v^*$. This happens for the remaining
$d^2$ contributions which read, after trivial manipulations,
\be
  \frac{-2\hbar d^2}{\pi}
  \tr\left[u\,\bar\gh O(1,1;\Lambda^2)\,\partial^3\, v^*
  \,\frac{\partial}{\bar\partial}\, \gh\veps^2\right].
\label{rel cont 0}
\ee

Writing out this functional trace in momentum space yields
\bea
  &\displaystyle{\frac{-2 i \hbar d^2}{\pi}
  \int\dif^2 x\,u(x) \int\frac{\dif^2 p}{(2\pi)^2}\,v^*(p)
  \int_0^1\dif t\,\frac{t}{(1+t)^4}}&
\nonumber\\
  &\displaystyle{ \times\left\{ \int\frac{\dif^2 k}{(2\pi)^2}\,
  \exp\left(-\frac{t}{1+t}\,\frac{(p+k)(\bar p+\bar k)}{\Lambda^2}
  - \frac{k\bar k}{\Lambda^2}\right)\,
  \frac{(p+k)^3 k}{\bar k} \right\}.}&
\nonumber
\eea
After shifting $k^\mu\,\rightarrow\,
\left(\sqrt{\frac{t+1}{2t+1}}\,k^\mu-\frac{t}{2t+1}\,p^\mu \right)$
to decouple the integrals over $k$ and $p$ we get
$$
  \frac{-2 i \hbar d^2}{\pi}
  \int\dif^2 x\,u(x) \int\frac{\dif^2 p}{(2\pi)^2}\,v^*(p)
  \int_0^1\dif t\,\frac{t}{(1+t)^3 (2t+1)}
  \exp\left(-\frac{t}{2t+1}\,\frac{p\bar p}{\Lambda^2}\right)
  \gh I(t,p;\Lambda^2),
$$
in terms of the momentum integral $\gh I(t,p;\Lambda^2)$
\be
  \gh I(t,p;\Lambda^2)= \int\frac{\dif^2 k}{(2\pi)^2}\,
  \exp\left(- \frac{k\bar k}{\Lambda^2}\right)
  \frac{ \left(\sqrt{\frac{t+1}{2t+1}}\,k-\frac{t}{2t+1}\,p \right)\,
   \left(\sqrt{\frac{t+1}{2t+1}}\,k+\frac{t+1}{2t+1}\,p \right)^3}
  {\left(\sqrt{\frac{t+1}{2t+1}}\,\bar k-\frac{t}{2t+1}\,\bar p \right)}.
\label{momint}
\ee
The net result of the presence of the operator $\bar\gh O(1,1;\Lambda^2)$
in \bref{rel cont 0} is the  $p$--dependent shift of the
integrand in \bref{momint}, without which the expression would vanish.
The angular integral can easily be done by converting it into a complex
contour integral, using $k = i \sqrt{\frac{x}{2}}\, |p|\, z$ and
$\bar k=-i\sqrt{\frac{x}{2}}\, |p|\, \frac1{z}$:
\bea
  \gh I(t,p;\Lambda^2)&=& -\frac{p^5}{(2\pi)^2}\,\frac{2t+1}{t}\,
  \int_{0}^{\infty} \dif x \,
  \exp{\left(-\frac{x}{2}\,\frac{|p|^2}{\Lambda^2}\right)}
\nonumber\\
  &&\times\oint_{S_1}{} \dif z
  \frac{\left(\sqrt x z \sqrt{\frac{t+1}{2t+1}}+\frac{t+1}{2t+1}\right)
  \,\left(\sqrt x z \sqrt{\frac{t+1}{2t+1}}-\frac{t}{2t+1}\right)}
  {\left(z -\sqrt x\, \frac{\sqrt{(t+1)(2t+1)}}{t}\right)},
\nonumber
\eea
where the contour integral in the complex plane is counterclock-wise over
the unit circle.
There is just one single pole, and only for a limited range of values
of $x$, which cuts off the $x$-integral to
$x\leq x_0(t)\equiv \frac{t^2}{(t+1)(2t+1)}$.
In terms of the rescaled variable $y=x/x_0(t)$ one finds
$$
  \gh I(t,p;\Lambda^2)= \frac{-(ip)^5}{2\pi}\,\frac{t^5}{(t+1)(2t+1)^4}
  \int_{0}^{1} \dif y \,
  \left( y+ \frac{t+1}{t} \right)^3 \left( y-1 \right)\,
  \exp{\left(-x_0(t)\,y\,\frac{p\bar p}{\Lambda^2}\right)}.
$$
At last the $\Lambda\rightarrow\infty$ limit can be taken.
The unique nonvanishing contribution to our expression
\bref{2loopw3bis} for the two--loop anomaly becomes
$$
\lim_{\Lambda^2\rightarrow\infty}\left\{
  \frac{-2 \hbar d^2}{\pi}
  \tr\left[u\,\bar\gh O(1,1;\Lambda^2)\,\partial^3\, v^*
  \,\frac{\partial}{\bar\partial} \gh\veps^2\right]\right\}=
  \frac{i \hbar d^2}{\pi^2}\,a\,
  \int\dif^2 x\,u\, \partial^5 \, v^*,
$$
with the numerical factor $a$ given by the integral
$$
  a= \int_0^1\dif t\, \frac{t^6}{(t+1)^4(2t+1)^5}
  \int_{0}^{1} \dif y \,
  \left( y+ \frac{t+1}{t} \right)^3 \left( y-1 \right)=
  \frac{1}{6!}.
$$

In summary, the expression for the two--loop anomaly \bref{2loopw3bis}
arising out of this computation is
\be
   \gh A_2(v^*, u)=\frac{i \hbar d^2}{720\pi^2}
   \int\dif^2 x \, u \,\partial^5 \,v^*.
\label{w3 twoloop}
\ee
This agrees with previous results in the literature,
obtained from operator product expansions \cite{hull91}, or
from the nonlocally regulated two-loop effective action \ct{p95}.

\subsection{Two--loop Consistency Condition}

\hspace{\parindent}%
With the explicit expressions \bref{w3 1loop} and \bref{w3 twoloop}
of the one and two loop anomalies for chiral $W_3$ gravity at hand, we
can finally pass to the verification of the two-loop consistency
condition \bref{cchla2}. The BRST variation of the two--loop anomaly is
\bea
   (\gh A_2,S)&=& \frac{id^2}{720\pi^2}\int\dif^2 x \left\{
   (\partial^5 v^*)[2(\partial c) u - c (\partial u)]+
   (\partial^5 u)[2 (\partial c) v^*-  c (\partial v^*)]\right\}
\nonumber\\
   &&+\frac{id^2}{720\pi^2}\int\dif^2 x
   (\partial^5 u)[ b^*(\partial u)-  2(\partial b^*) u].
\label{a2 s}
\eea
The fact that it does not vanish confirms,
notwithstanding ref.\,\ct{white92},
that higher loop consistency conditions in general do not have
the one--loop form \bref{oneloop}, but require the presence
of extra pieces related with the action of $\Delta$ on lower order
anomalies. In the absence of counterterms $M_p$, the computation
of such extra terms
using our general result eqs.\,\bref{omegaa1}--\bref{cchla2},
requires knowledge of the matrix of second derivatives
of the one--loop anomaly \bref{A1-5} with respect to
fields and antifields. The relevant nonvanishing entries
of $(\gh A_1)_{AB}$ are:
\bea
    (\gh A_1)_{ij}&=&
    -\frac{i}{6\pi}d_{ij}(\partial\, u\partial^3\, v^*\, \partial-
                          \partial\, v^*\,\partial^3\, u \,\partial)
\nonumber\\
    &&+\frac{ik}{2\pi}\delta_{ij}
    \left[\partial (u(\partial v^*) - v^*(\partial u)) \partial^3
    +\partial^3 (u(\partial v^*) - v^*(\partial u)) \partial\right]
\nonumber\\
    &&-\frac{ik}{6\pi}\delta_{ij}\partial
    \left[ 5 (\partial^3 u) v^* -12 (\partial^2 u)(\partial v^*)
    +12 (\partial u)(\partial^2 v^*) -5 u(\partial^3 v^*)\right]\partial,
\nonumber\\
    (\gh A_1)_{i b}&=& \frac{ik}{6\pi}d_i
    \left[\partial\, u\, \partial^3\, (u(\partial v^*) - v^*(\partial u))
    - \partial\, v^*\,\partial^3 (\partial u) u\right],
\nonumber\\
    (\gh A_1)_{i c}&=& \frac{i}{12\pi} d_i \partial\, v^*\,\partial^3,
    \quad\quad
    (\gh A_1)_{b c}= \frac{-iDk}{12\pi}
    (u(\partial v^*) - v^*(\partial u)) \partial^3,
\nonumber\\
    (\gh A_1)_{b u}&=& \frac{ik}{12\pi} \left[
    \partial^3 (D b^* -2\phi^* u+ 2v^*(\partial\phi))(u\partial)_{-1}
    -\partial^3(D c +2 u(\partial\phi))(v^*\partial)_{-1}\right.
\nonumber\\
      &&\left. \hspace{8mm}+2(\partial u) u\,\partial^3\,\phi^*
    -2 (u(\partial v^*)-v^*(\partial u)) \partial^3 (\partial\phi)\right],
\label{aij}
\eea
where $d_i$ stands for the contraction $d_i=\delta^{jk} d_{ijk}$
and $\phi$, $\phi^*$ for the field and antifield combinations
$\phi^i d_i$, $\phi^*_i d^i$, respectively. On the other hand, the
nonvanishing relevant entries for  $(\gh A_1)^A_{\,B}$ are:
\bea
    (\gh A_1)^i_{\,j}&=& -\frac{ik}{6\pi} \delta^i_j
    \left[ 6\partial\, u (\partial u) \partial^2
     +9 (\partial^2 u) (\partial u) \partial
     +8  u (\partial^3 u) \partial\right]+
     \frac{i}{6\pi} d^i_j u\, \partial^3\, u\, \partial,
\nonumber\\
    (\gh A_1)^i_{\,b}&=&
    -\frac{ik}{6\pi} d^i u\, \partial^3 (\partial u) u,
    \quad\quad
    (\gh A_1)^i_{\,c}= \frac{i}{12\pi} d^i\, u\, \partial^3,
\nonumber\\
    (\gh A_1)^{b^*}_{\,i}&=& -\frac{i}{12\pi} d_i \,\partial^3\, u\,
    \partial, \quad\quad
    (\gh A_1)^{b^*}_{\,b}=
    \frac{ikD}{12\pi} \partial^3 (\partial u) u,
\nonumber\\
    (\gh A_1)^{v^*}_{\,b}&=& \frac{ik}{12\pi}
    \left[(\partial\phi) \partial^3 (\partial u) u
    +2(\partial u)\partial^3 (D c+ 2 u (\partial\phi))
    +2\partial\, u(\partial^3 (D c+ 2 u (\partial\phi)))\right],
\nonumber\\
    (\gh A_1)^{c^*}_{\,i}&=&
    -\frac{ik}{6\pi} d_i u (\partial u) \partial^3\, u\, \partial,
    \quad\quad
    (\gh A_1)^{c^*}_{\,c}=
    -\frac{ikD}{12\pi} u (\partial u) \partial^3,
\nonumber\\
    (\gh A_1)^{c^*}_{\,u}&=&
    -\frac{ik}{12\pi}\left[ D (\partial^3 c)(u\partial)_{-1}
     + 2u (\partial u) \partial^3 (\partial\phi)
     + 2\partial^3\, u (\partial\phi) (u\partial)_{-1}\right].
\nonumber
\eea

The further computation of
$\Omega_{\gh A_1}$ \bref{omegaa1} is again organized by means of
an expansion in the number of antifields
\be
   \Omega_{\gh A_1}=
   (-1)^{(A+1)} \left\{ \left[ \gh\veps^2\, (\gh A_1)^A_{\,A}\right]
  +\left[\gh\veps^2\,\left((\gh A_1)^A_{\,B} \gh O^{BC}\,I_{CA}
   + S^A_{\,B}\gh O^{BC} (\gh A_1)_{CA}\right)\right]\right\}+
  O((\Phi^*)^2),
\label{antifield exp anom}
\ee
and as before only the finite terms linear in antifields
are really relevant for our purposes%
\footnote{ The antifield independent term in the expansion
           \bref{antifield exp anom} encodes potential divergencies.
           They are easily shown to vanish, using
           $\gh T(F,n)={\rm Tr}\left[\gh\veps^2(F\partial)_{n}\right]=0$
           \ct{p95}, which shows that no counterterms are necessary for
            this two--loop computation either.}.
Although there is a considerable number of terms, there is a very
limited number that have a non--zero limit.
With $\gh O$ as in \bref{ShadowPropW3} one obtains
\bea
   \lim_{\Lambda^2\rightarrow\infty} \Omega_{\gh A_1} &=&
   \frac{i d^2}{6\pi} \lim_{\Lambda^2\rightarrow\infty}
   {\rm Tr}\left\{\gh \veps^2
   \left[u\partial^3\,u\,\gh O\,\partial^2\,b^*\, \partial
   -c\,\gh O\,\partial^2\,
  (u\,\partial^3\,v^* \,\partial-
  v^*\,\partial^3\, u\, \partial)\right]\right\}
\nonumber\\
   &=& \frac{d^2}{720\pi^2}\int\dif^2 x
   \left[ 2 u (\partial^3 u)- 3(\partial u)(\partial^2 u)\right]
   (\partial^3 b^*)
\nonumber\\
   &+&\frac{d^2}{720\pi^2}\int\dif^2 x (\partial^2 c)
   \left[2(\partial^4 u) v^* - (\partial^3 u) (\partial v^*)
   + (\partial u) (\partial^3 v^*)- 2 u (\partial^4 v^*)\right].
\nonumber
\eea
An integration by parts and a glance
at the expression for the BRST variation of the two--loop anomaly
\bref{a2 s} confirms the correctness of
the two--loop consistency condition \bref{cchla2}.

We conclude our illustration of the nonlocally regularized computation of
higher loop anomalies and consistency conditions for chiral $W_3$
Gravity with some remarks. Previous computations of the higher loop
anomalies that appeared in the literature have largely rested,
explicitly or implicitly, on the assumed $W_3$ operator product expansion
of the quantum theory. For the $W_3$ case this results in a two--loop
anomaly. This was confirmed by a direct computation of the two--loop
effective action in \ct{p95} using nonlocal regularisation, and by a
regularised calculation of the anomaly itself in renormalised
Bogoliubov--Parasiuk--Hepp--Zimmerman perturbation theory in \ct{BPHZBV}.
The computation of this section uses, for the first time,
a regularized quantum master equation explicitly.
The related question, that of consistency conditions for such higher loop
anomalies, has been considered previously in ref.\,\ct{roya},
using the nonlocal consistency condition \bref{nlcc} rather than
its local consequence \bref{lcc}: it therefore completely differs
in spirit from ours, which is aimed at the {\em local} equations.
We have shown explicitly, in the two loop case, how to obtain these local
consistency conditions.

\newpage

\section{Conclusions}
\label{conclusions}

\hspace{\parindent}%
The question of the local anomalies emerges in the Field--Antifield or BV
formalism as a clash between the formal equations and the locality of the
theory, leading to the usual quantum field theory divergences. The problem
can be treated either by  eliminating the divergences as in the
treatment of \ct{BPHZBV} using
Bogoliubov--Parasiuk--Hepp--Zimmerman perturbation theory, or by
(temporarily) going to a nonlocal theory through a regularisation.
The regularisation in the Pauli--Villars spirit \ct{anombv}
is an example of this last method, but has the disadvantage that it
temporarily breaks the gauge invariance (which is afterwards restored
by finite counterterms) and that it is mainly suitable for one loop.
In this paper, a more drastic nonlocal regularisation scheme
\ct{emkw91} was
used, which works to all orders and respects gauge invariance,
even when considering non--trivial (open, reducible) gauge algebras since
it respects the BV structure \ct{p95}.

The nonlocality of this regularised theory completely eliminates the
divergence problem, so that the formal quantum BV master equation becomes
a sensible equation, giving an expression for anomalies (at arbitrary
order in $\hbar$),  and also consistency conditions that they must obey.
Removing the cutoff proved to be rather subtle however: it turned out that
some contributions to lower loop anomalies that vanish in this limit
nevertheless have quantum corrections that do not vanish. This phenomenon
was shown to be at the heart of the generation of higher loop anomalies in
this scheme.

We have used this mechanism to deduce an explicit expression for the
two--loop anomaly and its consistency condition, and shown the viability
of the scheme by explicitizing it for $W_3$ gravity. In particular it was
shown that the two loop anomaly satisfies a consistency condition that is
{\em different} from the celebrated equation of Wess and Zumino: it
contains an additional inhomogeneous term that is related to the presence
of a lower loop anomaly. It is obvious that the method extends
perturbatively to higher orders as well.

Since the consistency conditions in higher orders contain inhomogeneous
terms, it becomes a non--trivial task to characterise them
algebraically. For one loop, it is well known that they are characterised
by a local cohomology class. The specific regularisation that one choses
reflects on the specific representative of this class that comes out of
the actual computation, but one can shift between representatives by
adding local finite counterterms to the action. For two loops already
a complication arises. In the scheme adopted in this paper,
the consistency condition for the two--loop anomaly contains
the specific operator representing the one--loop anomaly,
in a form that is not obviously cohomological.
Stated differently, it depends explicitly on a possible counterterm
$M_1$, which may result in a renormalisation scheme dependence.
We leave the algebraic characterisation of these anomalies for the future.

\section*{Acknowledgements}

\hspace{\parindent}%
This work was carried out in the framework of the project
``Gauge theories, applied supersymmetry and quantum gravity'',
contract SC1--CT92--0789, of the European Economic Community.
J.\,P.\, acknowledges financial support from the spanish
ministry of education (MEC).

\endsecteqno

\end{document}